\documentclass[useAMS,usegraphicx,usenatbib]{mn2e}

\newcommand{\oii}{\mbox{${\rm [OII]\lambda3727}$}}
\newcommand{\oiiis}{\mbox{${\rm [OIII]\lambda5007}$}}
\newcommand{\oiii}{\mbox{${\rm [OIII]\lambda4959,\lambda5007}$}}
\newcommand{\oiiib}{\mbox{${\rm [OIII]\lambda4959}$}}
\newcommand{\oiiic}{\mbox{${\rm [OIII]\lambda4363}$}}
\newcommand{\sii}{\mbox{${\rm [SII]\lambda6717,\lambda6731}$}}
\newcommand{\halpha}{\mbox{${\rm H\alpha}$}}
\newcommand{\hbeta}{\mbox{${\rm H\beta}$}}
\newcommand{\rp}{\mbox{${\rm R_{23}}$}}
\newcommand{\doh}{\mbox{${\rm 12+\log(O/H)}$}}
\newcommand{\nii}{\mbox{${\rm [NII]\lambda6584}$}}
\newcommand{\oo}{\mbox{${\rm O_{32}}$}}
\newcommand{\vrot}{\mbox{${V_{\rm rot}}$}}      
\newcommand{\rdspec}{\mbox{${r_{\rm d,spec}}$}}   
\newcommand{\elfitpy}{\textsc{elfit2py}}
\newcommand{\elfitd}{\textsc{elfit2d}}
\newcommand{\iraf}{\textsc{iraf}}
\newcommand{\slantfrac}[2]{#1\!\left/#2\right.}
\newcommand{\snr}{$\slantfrac{S}{N}$}
\newcommand{\corr}{\Delta W^{+\rmn{abs}}_{\rmn{emi}}(\rmn{H}\beta)}
\newcommand{\unisim}{\sim\!}
\newcommand{\per}[1][1]{{$^{-#1}$}}

\title{The metallicities of luminous, massive field galaxies 
at intermediate redshifts.}

\author[Mouhcine et al.]{M. Mouhcine$^1$\thanks{Isaac Roberts Fellow}, 
S.P. Bamford$^2$\thanks{Current address: Institute of Cosmology and 
Gravitation, University of Portsmouth, Mercantile House, Hampshire Terrace,
Portsmouth, PO1 2EG, UK.}, A. Arag{\'o}n-Salamanca$^2$, O. Nakamura$^2$\\
$^{1}$Astrophysics Research Institute, Liverpool John Moores University,
Twelve Quays House, Egerton Wharf, Birkenhead, CH41 1LD, UK. \\
$^{1}$School of Physics and Astronomy, University of Nottingham,
       Nottingham, NG7 2RD, UK 
}

\date{Accepted ?.
      Received ?;
      in original form ?}
                                                    
\pagerange{\pageref{firstpage}--\pageref{lastpage}} \pubyear{2004}   
         
\begin{document}                                 
                                                          
\maketitle
                                                                     
\label{firstpage}                                        
  
\begin{abstract}

We derive oxygen abundances for a sample of $40$ luminous 
($M_{B} \la -19$), star-forming, mostly disk, field galaxies 
with redshifts in the range $0.2 \la z \la 0.8$, with a median 
of $\left< z \right> = 0.45$.
Oxygen abundances, relative to hydrogen, of the interstellar 
emitting gas are estimated by means of the empirically 
calibrated strong emission line ratio technique. The derived 
{\doh} values range from 8.4 to 9.0, with a median of 8.7. 
Twenty of these galaxies have securely measured rotation 
velocities, in the range 50--244 kms$^{-1}$.

The measured emission line equivalent widths and diagnostic 
ratios for the intermediate redshift galaxies cover similar 
ranges to those observed across a large sample of local 
galaxies. The estimated oxygen abundances for our luminous 
star-forming intermediate redshift galaxies cover the same
range as their local counterparts. However, at a given
galaxy luminosity, many of our galaxies have significantly 
lower oxygen abundances, i.e., ${\doh} \sim 8.6$, than local
galaxies with similar luminosities. Interestingly, these 
luminous, massive, intermediate redshift, star-forming galaxies 
with low oxygen abundances exhibit physical conditions, i.e., 
emission line equivalent width and ionization state, very
similar to those of local {\it faint} and metal-poor 
star-forming galaxies. The oxygen abundance of the interstellar 
gas does not seem to correlate with the maximum rotation 
velocity or the emission scale length of the parent galaxy. 
This suggests that there is a diversity in the intrinsic 
properties of the massive field galaxy population at 
intermediate redshifts 

The distribution of the colour excess, derived from the 
ratio of extinction-uncorrected {\hbeta} and {\oii} star 
formation rate indicators, covers a similar range to 
that observed locally, but exhibits a lower mean than is 
observed for local optically-selected star-forming galaxies.    
Luminous field galaxies at intermediate redshifts show 
similar star formation rates to their local counterparts. 
However, metal-poor, massive, star-forming galaxies tend to 
be systematically less affected by internal reddening than 
metal-rich, massive galaxies, which cover similar range 
of colour excess to local metal-rich luminous galaxies.  
Finally, the correlation between oxygen abundance and 
colour excess for intermediate redshift galaxies is found
to be similar to what is observed locally. This result 
indicates that the dust content of galaxies is more 
regulated by their chemical evolution rather than galaxy 
luminosity.

\end{abstract}                 


\section{Introduction}

The chemical abundances of stars and interstellar gas within 
galaxies provide a fundamental tool for tracing the evolution 
of galactic chemical enrichment.
Furthermore, the ratios of chemical element abundances help in 
constraining the star formation histories of galaxies. 
The determination of galactic chemical abundances at different
cosmic epochs can thus assist in constraining the likely 
scenarios of galaxy evolution.

With the advent of 10-m class telescopes and their powerful 
optical and near-infrared spectrographs, it is now possible 
to probe the properties of the interstellar gas in intermediate 
($0 < z < 1$; Kobulnicky \& Zaritsky 1999; Hammer et al. 2001; 
Lilly et al 2003; Liang et al. 2004, Maier et al. 2004, 
Kobulnicky \& Kewley 2004; Maier et al. 2005) and high-redshift 
($1.5 < z < 4$; Pettini et al. 1998, 2001; Kobulnicky \& Koo 
2000; Mehlert et al. 2002; Lemoine-Busserolle et al. 2003; 
Erb et al. 2003; Maier et al. 2006) galaxies. The classical 
nebular diagnostic techniques developed to study the properties 
of {H{\sc ii}} regions and emission line galaxies in the local 
universe are now used to study higher redshift galaxies (see 
Kobulnicky \& Zaritsky 1999).

The correlation between galaxy metallicity and luminosity 
in the local universe is one of the most significant 
observational results in galaxy chemical evolution studies 
(e.g., Lequeux et al. 1979; Skillman et al. 1989; 
Zaritsky et al. 1994; Melbourne \& Salzer 2002; Lamareille 
et al. 2004; Tremonti et al. 2004). Recent studies of galaxy 
evolution trace changes in 
scaling relations of galaxies as a function of cosmic epoch, 
such as the Tully--Fisher relation for disks (e.g., Simard 
\& Pritchet 1998; Ziegler et al. 2002; Milvang-Jensen et al. 
2003; Bamford et al. 2006; Nakamura et al. 2006) and the 
fundamental plane relation for spheroids (e.g., Im et al. 
2002; van Dokkum \& Ellis 2003). 
In this context, the evolution of the luminosity--metallicity 
relation of galaxies as a function of cosmic epochs can be 
used as a sensitive probe and consistency check of galaxy 
evolution.

The analysis of oxygen abundances of star-forming galaxies at 
intermediate redshifts seems to point toward the conclusion 
that the luminosity--metallicity relation evolves with 
redshift, with steeper slope at earlier cosmic time (Kobulnicky 
et al. 2003; Maier et al. 2004; Liang et al. 2004). 
Bright galaxies, with ${\it L\approx\,L_{*}}$, at intermediate 
redshift tend to have properties similar to bright galaxies in
the local universe, and fall on the local luminosity-metallicity 
relation.  However, faint galaxies seem to be systematically 
brighter for their metallicity in comparison to the local 
luminosity--metallicity relation. Lilly et al. (2003) found, 
however, that a significant fraction of a statistically complete 
sample of bright, i.e., $M_{B,AB} \le -20$, {\hbeta}-selected 
intermediate redshift galaxies appear to have low oxygen 
abundances, similar to what is seen in the local universe for 
galaxies with luminosities a factor of 10 or more lower. 
More recently, Maier et al. (2005) found that about one third 
of the Canada-France Redshift Survey galaxies they studied, 
with $M_{B,AB} \le -19.5$, have significantly lower oxygen 
abundances than do local galaxies with similar luminosities.

Intermediate redshift galaxies with low oxygen abundances have 
bluer colours than higher metallicity ones. Lilly et al. (2003) 
argue that these galaxies are immature massive galaxies that 
will increase their oxygen abundances to the present epoch in 
a downsizing scenario -- in which the manifestations of 
star-forming evolutionary activity appear in progressively more 
massive galaxies at earlier epochs -- rather than dwarf galaxies 
with their luminosities boosted by ongoing star formation 
events. The appearence of a population of massive galaxies with 
low oxygen abundances at $z \la 1.0$ is consistent with a later 
stellar mass assembly for a fraction of massive and 
intermediate-mass galaxies (Hammer et al. 2005).

The number of intermediate redshift galaxies with measured 
kinematics is relatively small. Bamford et al. (2005, 2006) and 
Nakamura et al. (2006) have extend the study of the kinematics 
of emission line galaxies at this redshift range by measuring 
the maximum rotation velocities for sizeable samples of luminous, 
i.e., $M_{B} \le -19.5$, cluster and field galaxies in the 
redshift range $0.1 \la z \la 1.0$ (see the quoted papers for 
more details). In the present paper, we analyze the properties 
of the interstellar emitting gas of field galaxies in the 
Bamford et al. (2005, 2006) and Nakamura et al. (2006) samples.
This will enable us to constrain the nature of the luminous 
galaxies with intermediate metallicities found at intermediate 
redshifts, and hence test the downsizing scenario for galaxy 
formation. The properties of cluster galaxies are analysed in 
another paper (Mouhcine et al. 2006).  

The paper is organized as follows. In Section~\ref{data} we 
describe the sample selection and emission line measurements. 
Distributions of galaxy properties are presented and 
discussed in Section~\ref{spec}. In Section~\ref{prop}, we 
present the luminosity--metallicity relation, star formation 
rates, and interstellar extinction of our galaxy sample.
The implications of our results are discussed in 
Section~\ref{disc}, and we summarize our conclusions in 
Section~\ref{concl}.

A concordance cosmological model with ${\rm 
H_{\circ}=70 km\,s^{-1}}$, $\Omega_{\Lambda}=0.7$, 
$\Omega_{m}=0.3$ has been adopted throughout the paper.

\section{Observations and sample selection}
\label{data}

The target selection, photometry and spectroscopic data 
reduction have been described thoroughly elsewhere.
Bamford et al. (2005) covers most of the VLT observations, 
Milvang-Jensen et al. (2003), Milvang-Jensen (2003) detail 
the VLT MS1054 data, and Nakamura et al. (2006) describes 
the Subaru observations. We therefore only briefly summarize 
this information here, before describing in more detail the 
equivalent width measurements utilized in this study.

The observed galaxies are located in nine fields, centred 
on galaxy clusters at redshifts in the range 
$z=0.20$--$0.83$. Spectroscopy and imaging were obtained 
using FORS2\footnote{http://www.eso.org/instruments/fors} 
on the VLT (Seifert et al. 2000), and 
FOCAS\footnote{http://www.naoj.org/Observing/Instruments/FOCAS}
on the Subaru telescope (Kashikawa et al. 2002). 
The galaxies observed within each field were selected by 
assigning priorities based upon the likelihood of being 
able to measure a rotation curve.  Higher priorities were 
assigned for each of the following: disky morphology, 
favourable inclination for rotation velocity measurement, 
known emission line spectrum, and available Hubble Space
Telescope data. It is important to note that this priority 
ranking method preferentially selects bright, star-forming 
disc galaxies, and therefore we are not probing the average 
galaxy population.  However, this sample is still useful 
in examining the evolution of the brightest star-forming 
disc galaxies (see Bamford et al. 2005). 

The absolute rest-frame $B$-band magnitudes utilised in this 
study are derived from a collection of multi-band imaging.  
For each galaxy, the available observed apparent magnitudes 
were converted to apparent rest-frame $B$-band using colour- 
and $k$- corrections determined from the spectral energy 
distribution best-fitting the colour data. The bands 
available for each galaxy vary, and are listed in Table A1 
of Bamford et al. (2005), along with the band upon which the 
rest-frame $B$-band magnitude is based. This is always the 
closest observed match to rest-frame $B$ available.  
The magnitudes are corrected for Galactic extinction using
the maps and conversions of Schlegel, Finkbeiner \& Davis 
(1998), and for internal extinction (including face-on 
extinction of $0.27$ mag), following the prescription of 
Tully \& Fouque (1985). Note that variations in internal 
extinction levels, as investigated later in this paper, 
likely produce both random and systematic uncertainties 
of $\sim 0.1$--$0.2$ mag on this correction. 
However, these errors are comparable to those introduced 
through uncertainties in the inclination measurement, and 
in any case are generally smaller than the differences found 
below between our intermediate-redshift and local comparison 
samples. All magnitudes given in this paper are on the Vega 
zeropoint system.  For further details see Bamford et al. 
(2005).

From the reduced two-dimensional spectra, individual sky- 
and continuum-subtracted emission line `postage stamps' were 
extracted.  In order to measure the rotation velocity ($\vrot$) 
and emission scale lengths ($\rdspec$) each emission line was 
fitted independently using a program named \elfitpy, which is
based on the algorithms of \elfitd{} by Simard \& Pritchet
(1998, 1999). In order to produce a single value of $\vrot$ 
and $\rdspec$ for each galaxy the values for the individual 
lines are combined by a weighted mean. The final galaxy 
kinematic sample contains 145 galaxies, including both field 
and cluster galaxies.

In order to measure line equivalent widths and fluxes the 
two-dimensional spectra were averaged in the spatial direction 
to produce one-dimensional spectra.  The spatial region used 
for each spectrum was determined by averaging the two-dimentional 
spectrum in the wavelength direction, fitting a symmetrical 
(Voigt) profile, and calculating the distance from the fit 
centre to where the profile falls to approximately 1\% of it's 
peak value.  All lines were fit with a combination of Gaussians 
using the iterative \textsc{amoeba} minimization algorithm of 
the \iraf/\textsc{stsdas} task \textsc{ngaussfit}. 
Once the best-fitting parameters (i.e. continuum level, position, 
amplitude and full width at half maximum for a single line) are 
found, their uncertainties are estimated through a Monte-Carlo 
resampling technique. In this, twenty synthetic data realisations 
are created from the best-fitting function, with random noise 
added to each pixel corresponding to the provided error image.  
These synthetic spectra are then fit in the same manner as the
true data, and the standard deviations of the resultant 
parameters are used to estimate the uncertainties on the 
best-fitting parameters for the true data.

Single Gaussian fits were attempted for all visible emission 
lines. Typically studies account for the effect of the underlying 
stellar Balmer lines in absorption by applying a correction to 
the rest-frame equivalent width of the \hbeta\ emission line 
that is independent of the galaxy properties. 
Different correction values for {\hbeta} stellar absorption 
have been used in the literature, ranging from 2\AA\ (e.g., 
Kobulnicky et al. 2003) to 5\AA\ (e.g., Kennicutt 1992).

\begin{figure}
\includegraphics[clip=,width=0.45\textwidth,angle=270]
{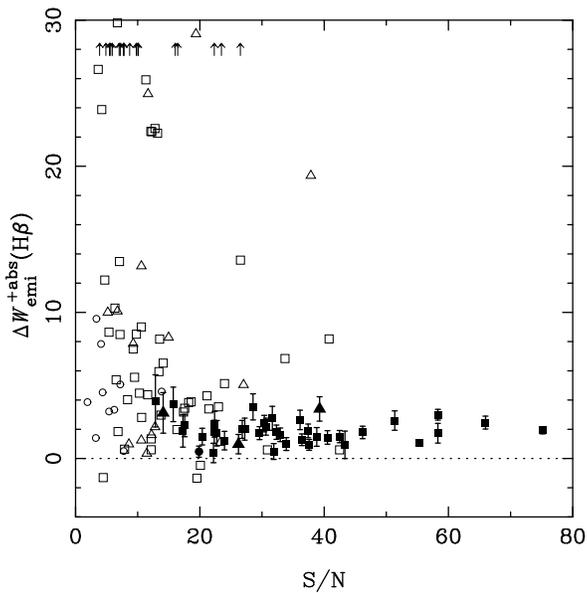}
\caption{The {\hbeta} absorption correction, $\corr$, 
determined from the difference between the emission line 
equivalent width measurements of two-component and
one-component fits, plotted versus the signal-to-noise 
in the adjacent continuum, \snr.  The filled points are 
those which have robust two-component fits, as judged by 
the cuts described in the text and indicated in 
Figure \ref{fig:corr_selection}.  Error bars are omitted 
from the open points for clarity.}
\label{fig:corr_sn}
\end{figure}

Balmer emission lines in the spectra of our galaxy sample 
have been corrected from the underlying stellar absorption 
by considering simultaneous fits of the emission and absorption 
lines. The two-component fits to the Balmer lines were found to 
be unreliable for a fraction of galaxy spectra, particularly 
for low \snr{} data. Figure \ref{fig:corr_sn} shows the 
difference in the emission line equivalent width measured 
from the two- and one-component fits ($W^{(2)}_{\rmn{emi}}$ 
and $W^{(1)}_{\rmn{emi}}$ respectively). This quantity is 
effectively the correction, $\corr$, one would need to apply 
to the one-component fit emission line equivalent widths to 
obtain the true equivalent width in the absence of stellar 
absorption,
\begin{equation}
\corr = W^{(2)}_{\rmn{emi}} - W^{(1)}_{\rmn{emi}} \; .
\end{equation}

\begin{figure}
\includegraphics[clip=,width=0.45\textwidth,angle=270]
{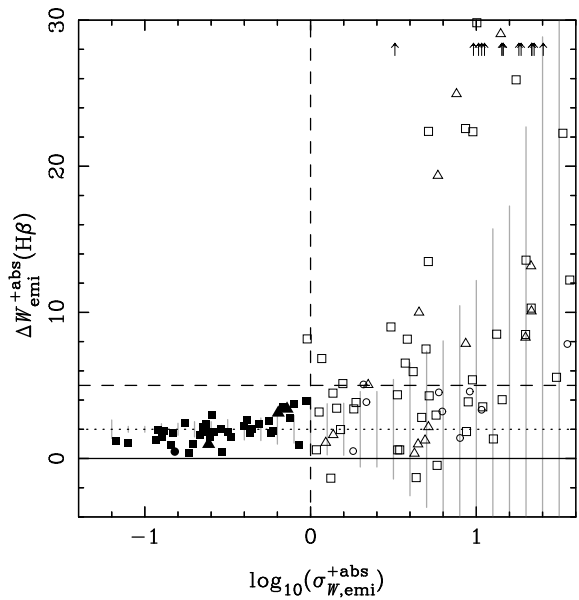}
\caption{The H$\beta$ absorption correction, $\corr$, 
determined from the difference between the emission line 
equivalent width measurements of two-component and
one-component fits, plotted versus an estimate the 
additional error introduced into the EW measurement by 
the inclusion of the absorption component, 
$\sigma^{+\rmn{abs}}_\rmn{EW,emi}$. Measurements from the 
2002 VLT data are plotted with squares, the 2001 VLT data 
are plotted with circles, and Subaru data are plotted using 
triangles. Dashed lines show the cuts adopted in selecting 
the reliable sample of two-component fits. Those points 
which pass these cuts are filled, while those rejected are 
open. The mean $\corr$ of the reliable points, used to 
correct the one-component H$\beta$ fits for the effects of 
stellar absorption, is indicated by the dotted line.
The gray vertical lines indicate the average error on the 
data points, in bins of $0.1$ dex in 
$\sigma^{+\rmn{abs}}_\rmn{EW,emi}$.}
\label{fig:corr_selection}
\end{figure}

The one-component fits should underestimate the equivalent 
width in cases where Balmer line absorption is additionally 
present. The correction introduced by including this effect, 
namely through a two-component fit, typically amounts to 
$\unisim 2$\AA{}. It can be seen that there is 
a ridge around $\corr$ at high \snr, and very large scatter 
at low \snr. However, highly deviant points continue to be 
present upto fairly high \snr.  Virtually all of these 
outlying points are due to a failure of the two-component 
fit. The nature of the two-component fit makes it prone 
to fitting noise features either side of the emission line 
using a very deep absorption line, and hence requiring an 
overly large emission component to still fit the central 
emission line. However, when attempting to fit noise 
features the best-fitting parameters naturally have high 
uncertainties. In order to identify unreliable two-component 
fits we estimate the additional error introduced into the 
equivalent width measurement by the inclusion of the absorption 
component as
\begin{equation}
\sigma^{+\rmn{abs}}_{W,\rmn{emi}} = 
\sqrt{{\sigma^{(2)}_{W,\rmn{emi}}}^{\!\!\!\!2} - 
           {\sigma^{(1)}_{W,\rmn{emi}}}^{\!\!\!\!2}\:} \; .
\end{equation}
A plot of $\corr$ versus $\sigma^{+\rmn{abs}}_{W,\rmn{emi}}$ 
does a considerably better job of separating points with 
reasonable $\corr$ values from those with unrealistically high 
$\corr$, as can be seen in Figure \ref{fig:corr_selection}.

From examining Figure \ref{fig:corr_selection}, it is clear 
that points with $\sigma^{+\rmn{abs}}_\rmn{EW,emi} < 1.0$\AA{} 
nearly all lie around the value of $\unisim 2$\AA.  
For $\sigma^{+\rmn{abs}}_\rmn{EW,emi}$ above this value there 
is a striking increase in both the scatter, and the number of 
unrealistically discrepant points. Note that
$\sigma^{+\rmn{abs}}_\rmn{EW,emi} < 1.0$\AA{} corresponds 
to where the uncertainty contributed by the inclusion of an 
absorption component in the fit is less than half the expected 
size of the correction.  That is, at 
$\sigma^{+\rmn{abs}}_\rmn{EW,emi} > 1.0$\AA{} we would expect 
$>10$\% of points to lie further from the true value of the 
correction than the size of the correction itself, simply due 
to the measurement uncertainty. It therefore seems reasonable 
to only consider two-component fits for which
$\sigma^{+\rmn{abs}}_\rmn{EW,emi} < 1.0$\AA{}.  In addition, 
a cut on $\corr$ is also useful in order to discard clearly 
discrepant points with $\corr > 5$\AA. These cuts are indicated 
in Figure \ref{fig:corr_selection}.

The mean of $\corr$ for spectra with both reliable two- and 
one-component fits to H$\beta$ is $1.9$\AA, with scatter 
$0.9$\AA. Accounting for measurement uncertainties leaves 
an intrinsic scatter about the mean of $\unisim 0.6$\AA.
For galaxies with reliable two-component Balmer fits we use 
the emission equivalent width determined from this fit.  
For all other galaxies we use the result of the one-component 
fit, and apply a uniform correction for the effect of stellar 
absorption. For this correction we adopt the mean of that 
determined for the two-component fits,
 $\left<\corr\right> = 1.9$\AA, and include an additional 
uncertainty of $0.6$\AA{} in quaderature.  

Our entire sample of field galaxies with identifiable emission 
lines contains $212$ galaxies, and spans a redshift range of 
$0.04$ to $1.22$ with a median of $0.36$. We searched within 
this spectroscopic sample for galaxies with emission lines 
suitable for chemical analysis. Only galaxies for which it 
was possible to measure {\oii}, {\hbeta}, and {\oiiis} emission 
lines were retained, i.e., the lines needed to determine the 
ionizing source and to measure the oxygen abundance. 
After applying these selection criteria the sample size drops 
to $60$. For reliable oxygen abundance determinations, only 
galaxies for which {\hbeta} emission line is well detected are 
retained.  This was judged by requiring the {\snr}, estimated 
from the median pixel value in regions $29$--$58$\AA{} away 
from the line on both sides, divided by the median value of 
the error image in the same region, to be larger than 8. 
An additional $16$ objects were excluded from the sample due 
to the weak detection of {\hbeta}. Of the the original data 
set with $212$ objects, the final sample contains $44$ emission 
line galaxies. Twenty-one of the sample galaxies have securely 
measured rotation velocities.\footnote{These numbers reduce to
$40$ and $20$ respectively once four galaxies are rejected due
to difficulties in measuring their metallicities;
see section \ref{oxy_abund}.}

\section{Spectral analysis}
\label{spec}

\subsection{Contamination from active galactic nuclei }

Nonthermal sources, such as active galactic nuclei (AGN), 
often produce emission-line spectra that superficially resemble 
those of star-forming regions. AGN must be identified as such 
because blindly applying emission-line metallicity diagnostics 
calibrated from {H{\sc ii}} region photoionization models will 
produce erroneous metallicities.

Traditionally, AGN can be distinguished from star-forming 
galaxies using the classical diagnostic ratios of two pairs 
of relatively strong emission lines, i.e., {\oiiis}/{\hbeta}
vs. {\nii}/{\halpha}, and/or {\oiiis}/{\hbeta} vs. 
{\sii}/{\halpha}  diagrams (Baldwin et al. 1981, Veilleux 
\& Osterbrock 1987). Some of these emission lines are usually 
not available for galaxies at $z \ga 0.3$. To solve this 
Rola et al. (1997) have investigated the location of starbursts 
and AGNs in diagnostic diagrams involving {\it blue} emission 
lines only, i.e. {\oii}, {\hbeta}, and {\oiiib} (see also 
Dessauges-Zavadsky et al. 2000). More recently, 
Lamareille et al. (2004) have used a large sample of emission 
line galaxies drawn from the Two Degree Field Galaxy Redshift 
Survey to investigate the separation between starbursts and 
AGNs using equivalent width ratios of blue emission line ratios 
as diagnostics for the ionizing source. 

None of the observed spectra contain all the required emission 
lines to determine the ionizing source using the classical 
techniques. Thus, we use the equivalent width ratios of 
{\oiii}/{\hbeta} and {\oii}/{\hbeta} as parametrized by 
Lamareille et al. (2004) to check for the presence of AGNs 
in our sample. Fig.~\ref{diagn_diagr} shows the distribution 
of the sample of intermediate redshift field galaxies in this
diagnostic diagram. The solid curve shows the demarcation 
line, separating starburst galaxies from AGN, of Lamareille 
et al. (2004). All the objects in our sample fall within 
the zone where starburst galaxies are located, indicating
that in all of them the source ionizing the interstellar
gas is an episode of star formation. Table~\ref{gal_coord} 
lists the coordinates of the final sample of field galaxies.

\subsection{Basic sample properties}

Table~\ref{gal_prop} lists the redshifts, absolute B-band 
magnitudes, {\oii}, {\hbeta}, and {\oiiis} emission line 
equivalent widths, maximum rotation velocities, and 
emission scale length for the objects in our final sample.  

\begin{figure}
\includegraphics[clip=,width=0.45\textwidth]{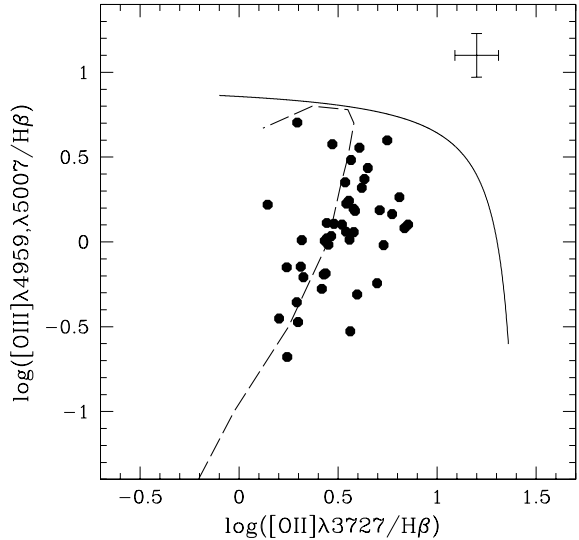}
\caption{Diagnostic diagram for our sample of intermediate 
redshift emission line galaxies. The solid line shows 
the proposed separation between starburst galaxies and AGNs 
by Lamareille et al. (2004). The dashed line indicates the 
theoretical sequence of McCall et al. (1985), which fits
the local {H{\sc ii}} regions with metallicity increasing
from right to left.}
\label{diagn_diagr}
\end{figure}

The dashed line in Fig.~\ref{diagn_diagr} shows the theoretical 
sequence of McCall, Rybski, \& Shields (1985) for line ratios 
of {H{\sc ii}} galaxies as a function of metallicity, which 
nicely fits measurements of local {H{\sc ii}} galaxies. 
The track is interpreted as a metallicity-excitation sequence; 
along the track the metallicity is high at the lower left, 
i.e., for low excitation systems, and low at the upper right, 
i.e., for high excitation systems (McCall et al. 1985). 
Intermediate redshift emission-line galaxies define a 
continuous sequence, strikingly consistent with that defined 
by local {H{\sc ii}} galaxies. Our sample contains objects 
with a wide variety of excitation levels, extending from 
those observed for local luminous galaxies to levels observed 
for faint and metal-poor dwarf galaxies at the present epoch. 
This might suggest that our sample of intermediate redshfit, 
luminous, star-forming field galaxies contains both 
low-metallicity and metal-rich systems. 

\begin{figure*}
\includegraphics[clip=,width=0.22\textwidth]{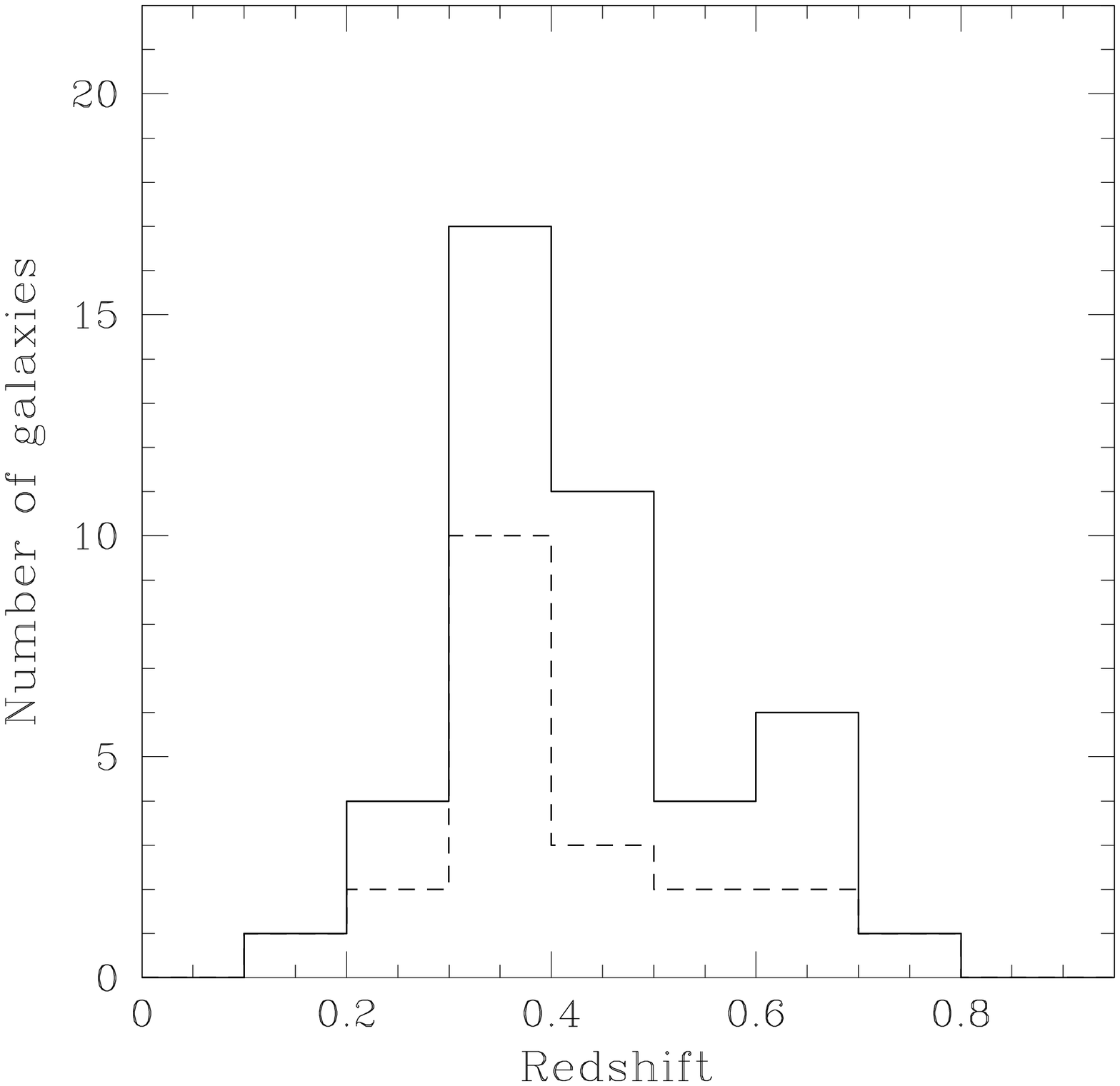}
\includegraphics[clip=,width=0.22\textwidth]{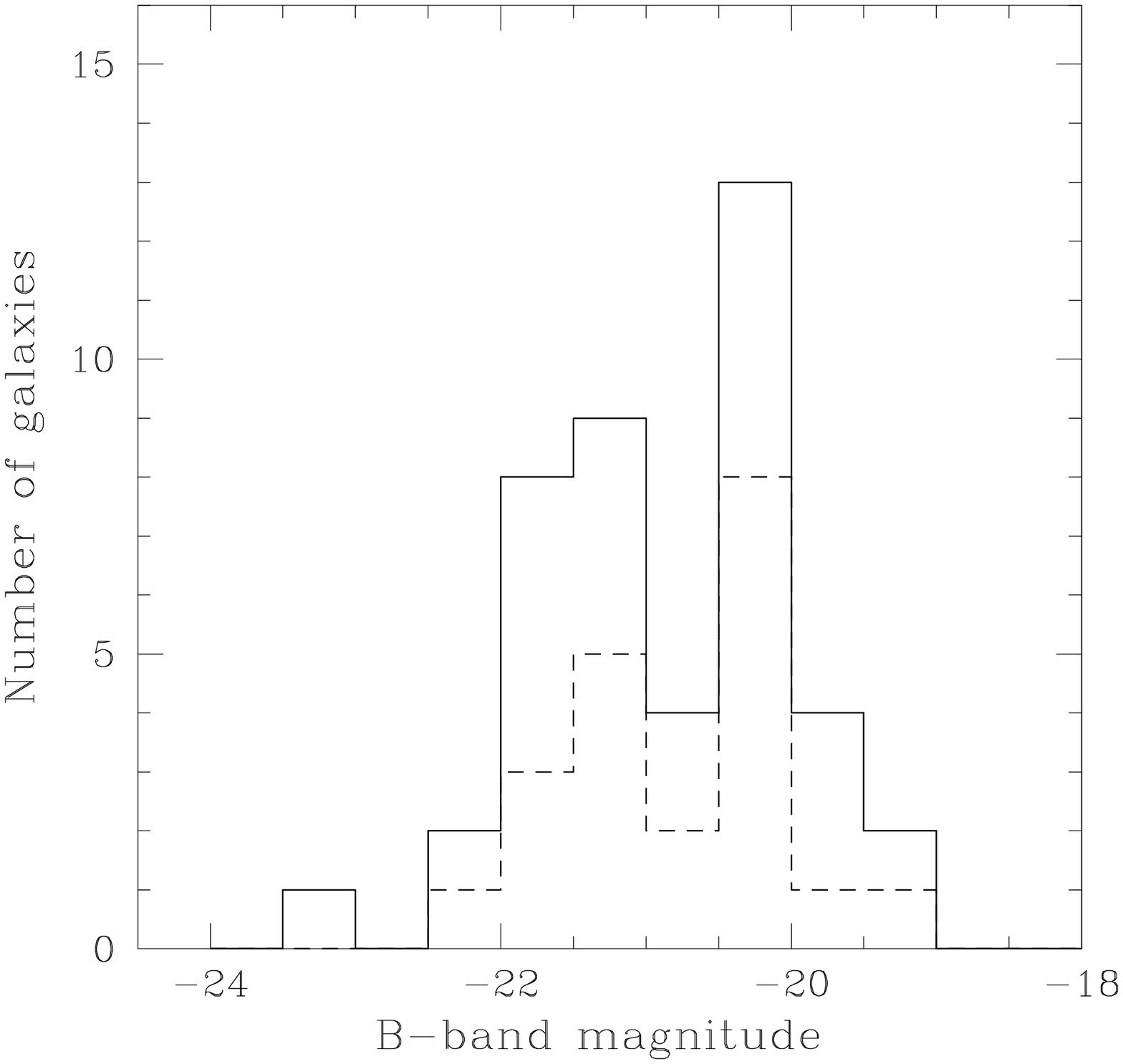}
\includegraphics[clip=,width=0.22\textwidth]{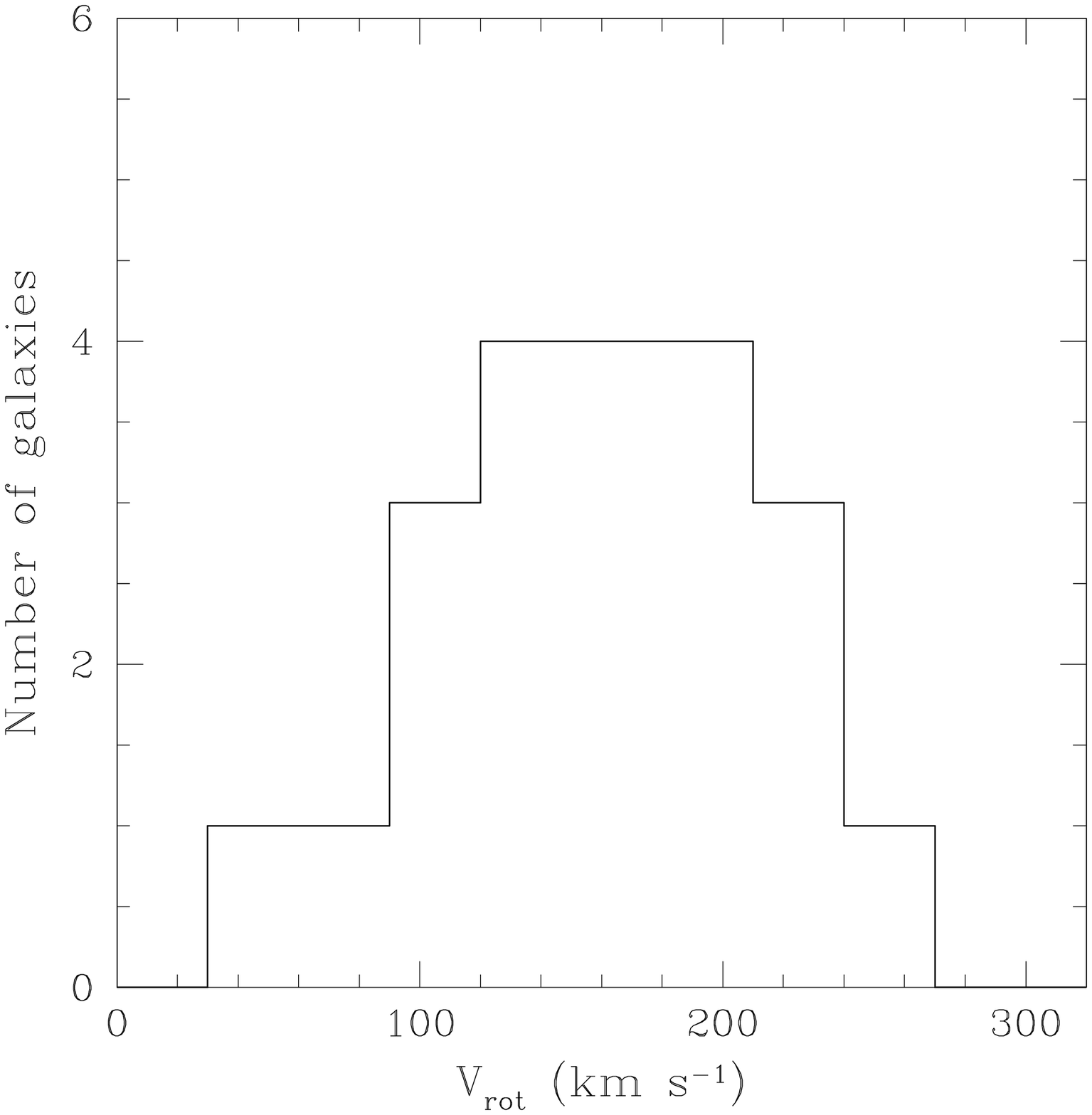}
\includegraphics[clip=,width=0.22\textwidth]{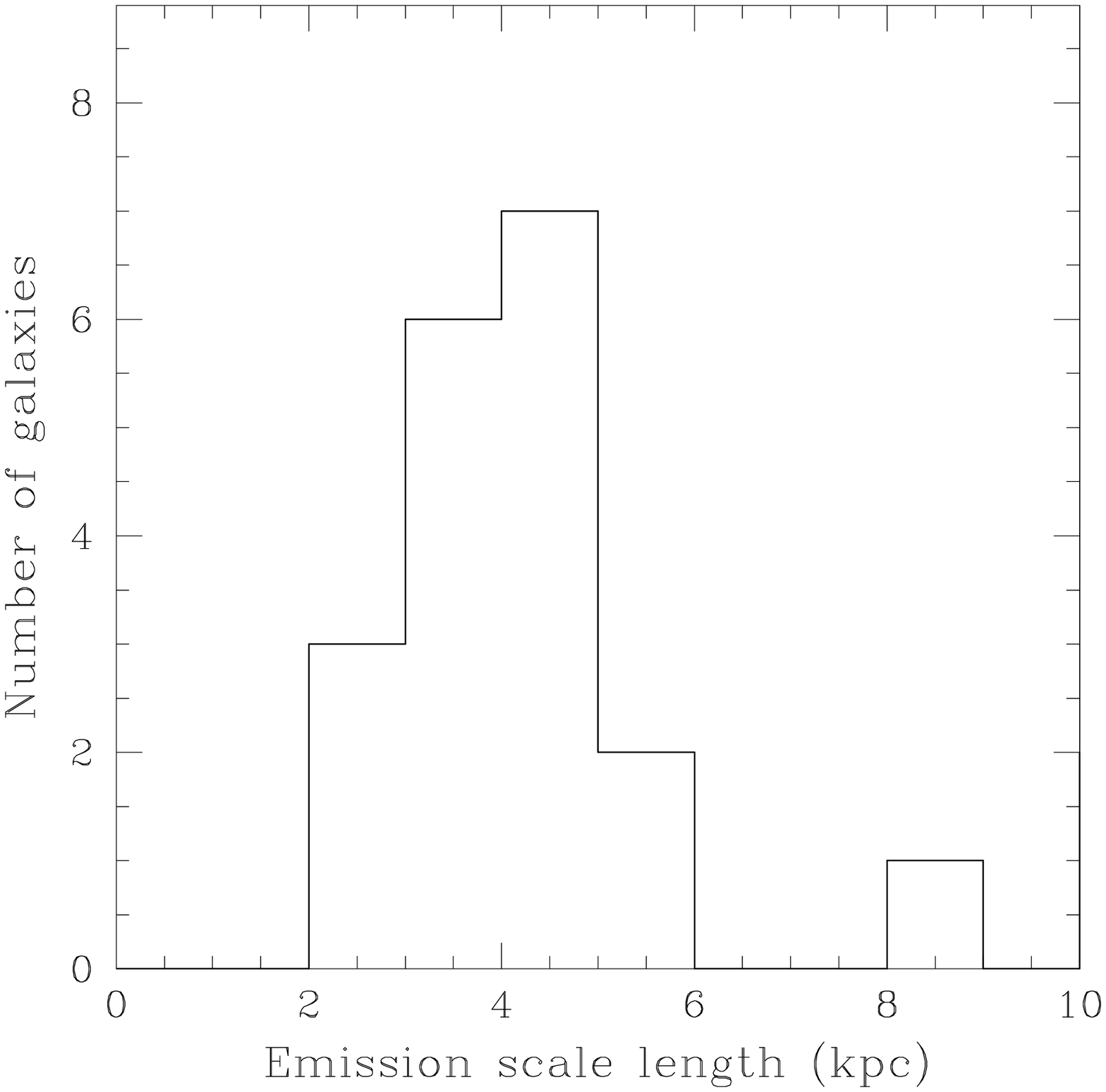}
\includegraphics[clip=,width=0.22\textwidth]{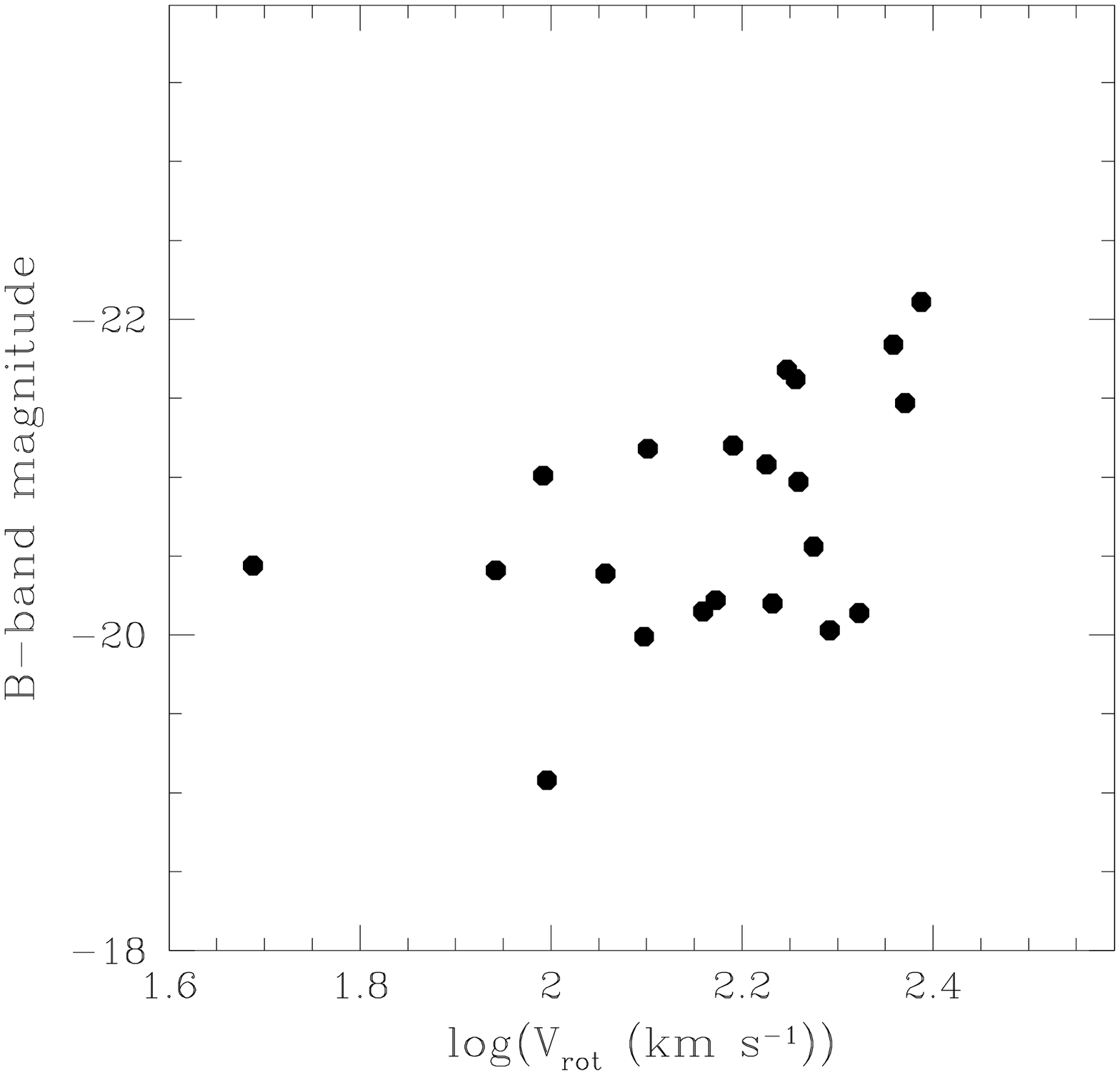}
\includegraphics[clip=,width=0.22\textwidth]{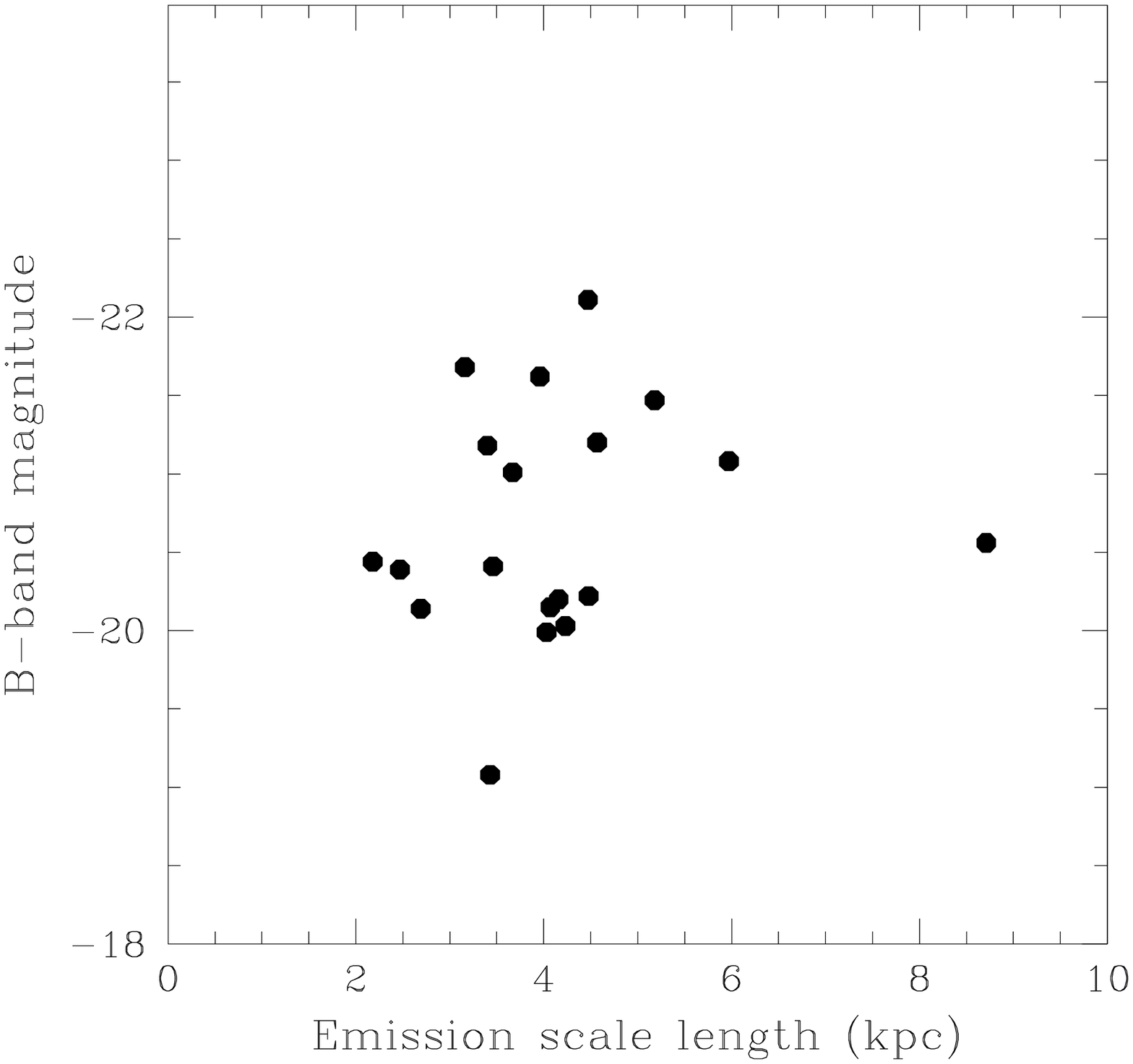}
\includegraphics[clip=,width=0.22\textwidth]{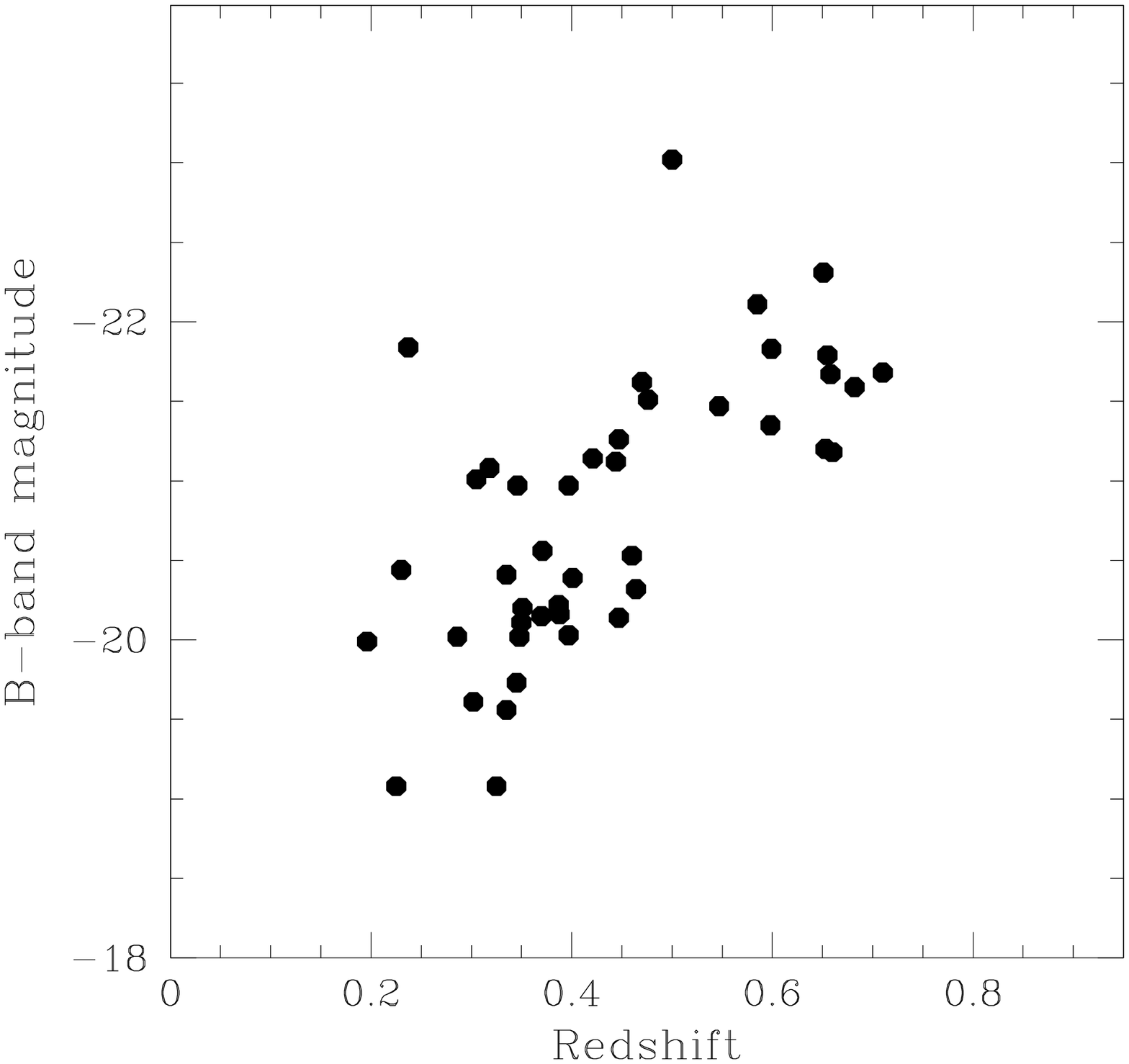}
\includegraphics[clip=,width=0.22\textwidth]{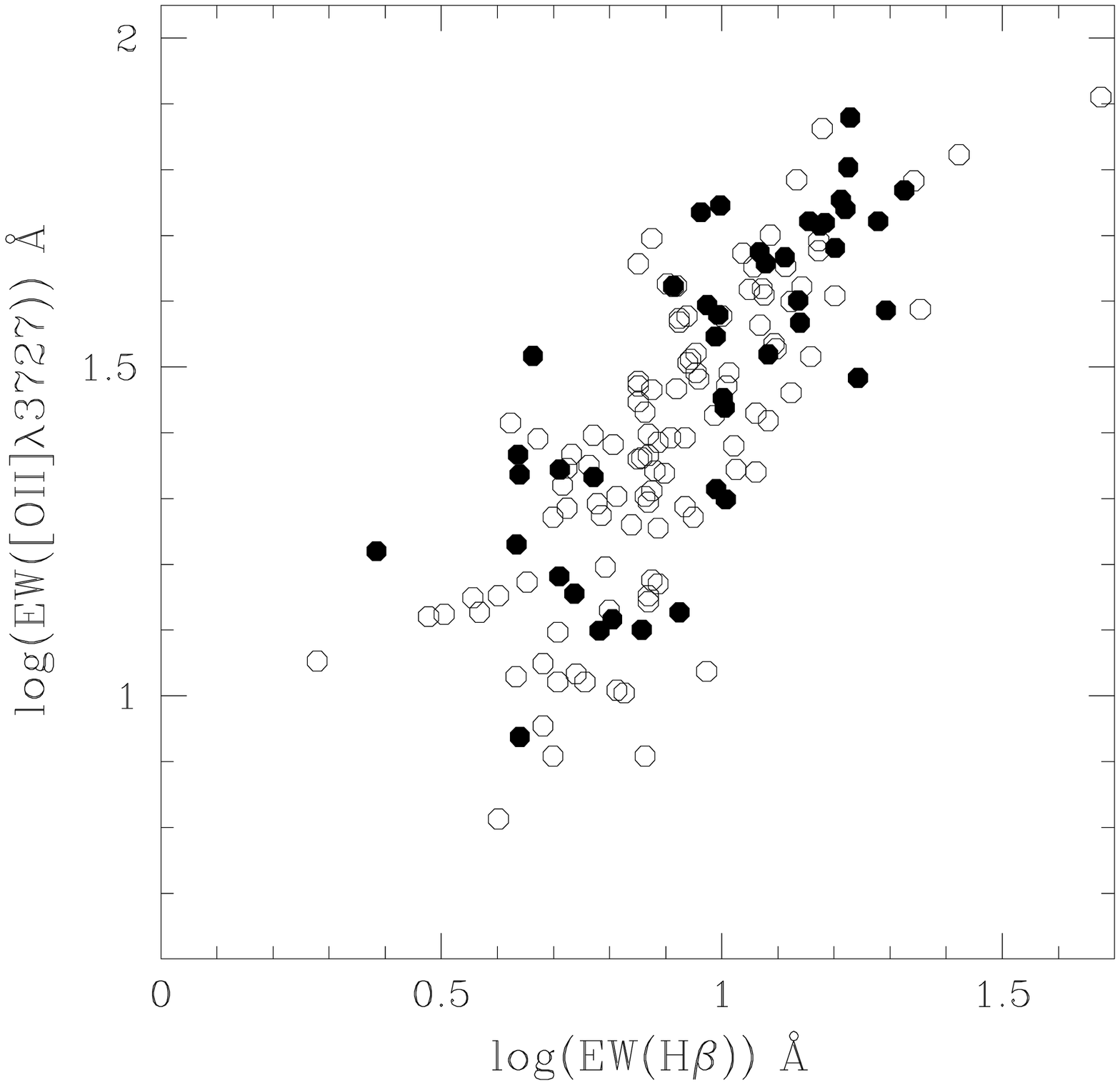}
\caption{Upper panels: distribution of redshift, B-band 
absolute magnitude, rotation velocity, and emission 
length scale for our final sample of star-forming galaxies. 
Dashed histograms show the distribution of galaxies with 
measured maximum rotation velocity. Bottom panels: plots 
of maximum rotation velocity, emission scale length, and
redshift versus B-band magnitude.
The right panel shows the correlation between the rest
frame equivalent widths of {\oii} and {\hbeta} emission
lines for our intermediate redshift galaxy sample, shown 
as filled circles, and a local comparative sample from 
Jansen et al (2000); see text for more details.}
\label{sample_prop}
\end{figure*}

The upper panels of Fig.~\ref{sample_prop} show the 
distribution of redshift, B-band absolute magnitude, 
rotation velocity, and emission scale length for the 
final sample of intermediate redshift star-forming 
galaxies. The dashed histograms show the distribution
of redshift and B-band absolute magnitude for galaxies 
with measured rotation velocities. Our sample of 
star-forming galaxies at intermediate redshifts is 
dominated by bright ($M_{B} \la -19$), massive 
($\vrot \ga {\rm 80\,km\,s^{-1}}$), and large disk
galaxies ($\rdspec \ga {\rm 2\,kpc}$).
The lower panels of Fig.~\ref{sample_prop} show the 
Tully--Fisher relation and the galaxy magnitude--size 
relation for our sample (see Bamford et al. 2005, 2006 
and Nakamura et al. 2006 for more details). The redshift 
versus galaxy luminosity relation displays a correlation, 
i.e., only high luminosity galaxies are observed at 
higher redshift, while low-luminosity galaxies are 
detected mostly at low redshifts. Our galaxy sample is 
therefore biased toward brighter galaxies at the highest 
end of the redshift distribution. This is a type of 
selection bias expected for flux-limited spectroscopic 
surveys. 
The bottom right panel of Fig.\,\ref{sample_prop} shows
the relationship between the rest-frame equivalent widths
of {\oii} and {\hbeta} emission lines for our sample of 
intermediate redshift star-forming galaxies, shown as 
filled circles, and a comparative local sample of star-forming
galaxies, shown as open circles (see Section~\ref{diag_sect}
for more details). This figure shows that the overall 
distribution of {\oii} rest-frame equivalent width for 
massive and luminous star-forming galaxies at intermediate 
redshifts is similar to the distribution for local 
star-forming galaxies.

\subsection{Diagnostic diagrams}
\label{diag_sect}

Different emission line ratios are sensitive to the
metallicity and the level of ionization of the emitting 
gas, and may be used to probe the physical conditions 
of interstellar star-forming gas.

As a local galaxy sample to compare the properties 
of intermediate redshift galaxies with, and thus to 
investigate variation in the properties of 
star-forming galaxies between $z \sim 0.5$--$1.0$ and 
the present epoch, we use the sample of Jansen et al. 
(2000), who observed the Nearby Field Galaxy Sample 
(NFGS) of about 200 galaxies. The galaxy sample was 
selected from the first CfA redshift catalog 
(Huchra et al. 1983) to approximate the local galaxy 
luminosity function.
The NFGS sample includes galaxies of all morphological 
types and spans 8 mag in luminosity and a broad range 
of environments. The spectra of NFGS sample galaxies 
are integrated over a significant fraction of the 
galaxies, and are thus similar to the unresolved spectra 
of the intermediate redshift galaxies in our sample. 
Following Jansen et al. (2000), an additional correction 
for stellar absorption of 1.5 {\AA} (1 {\AA}) has been 
applied to the NFGS {\halpha} ({\hbeta}) equivalent 
widths.

We use the classical diagnostic ratios of two pairs of 
relatively strong emission lines (Baldwin et al. 1981; 
Veilleux \& Osterbrock 1987) to distinguish between 
galaxies dominated by emission from star-forming regions 
and galaxies dominated by emission from non-thermal 
ionizing sources. 
We classify galaxies according to their position in 
{\oiiis}/{\hbeta} vs. {\nii}/{\halpha} and {\oiiis}/{\hbeta} 
vs. {\sii}/{\halpha} diagrams. The demarcation between 
star-forming galaxies and AGNs in both diagrams was taken 
from Kewley et al. (2001). We used the conservative 
requirement that a galaxy must be classified as a 
star-forming galaxy in both diagnostic diagrams in order 
to be retained in the local conparison sample of star-forming 
galaxies. 

The colour excess from obscuration by dust for NFGS sample 
galaxies was estimated from the observed ratio of {\halpha} 
and {\hbeta} line fluxes. We adopt the Milky Way interstellar 
extinction law of Cardelli, Clayton, \& Mathis (1989), with 
$R_{V}=3.1$. We assume an intrinsic ratio Balmer decrement 
of $2.85$, corresponding to the case B recombination with 
a temperature of ${\rm T=10^4 K}$ and a density of ${\rm 
n_{e}\sim\,10^2-10^4\,cm^{-2}}$ (Osterbrock 1989). 
The published extinction laws are similar in the optical, 
making the determination of the colour excess independent 
of the exact choice of the extinction law.

\begin{figure*} 
\includegraphics[clip=,width=0.45\textwidth]{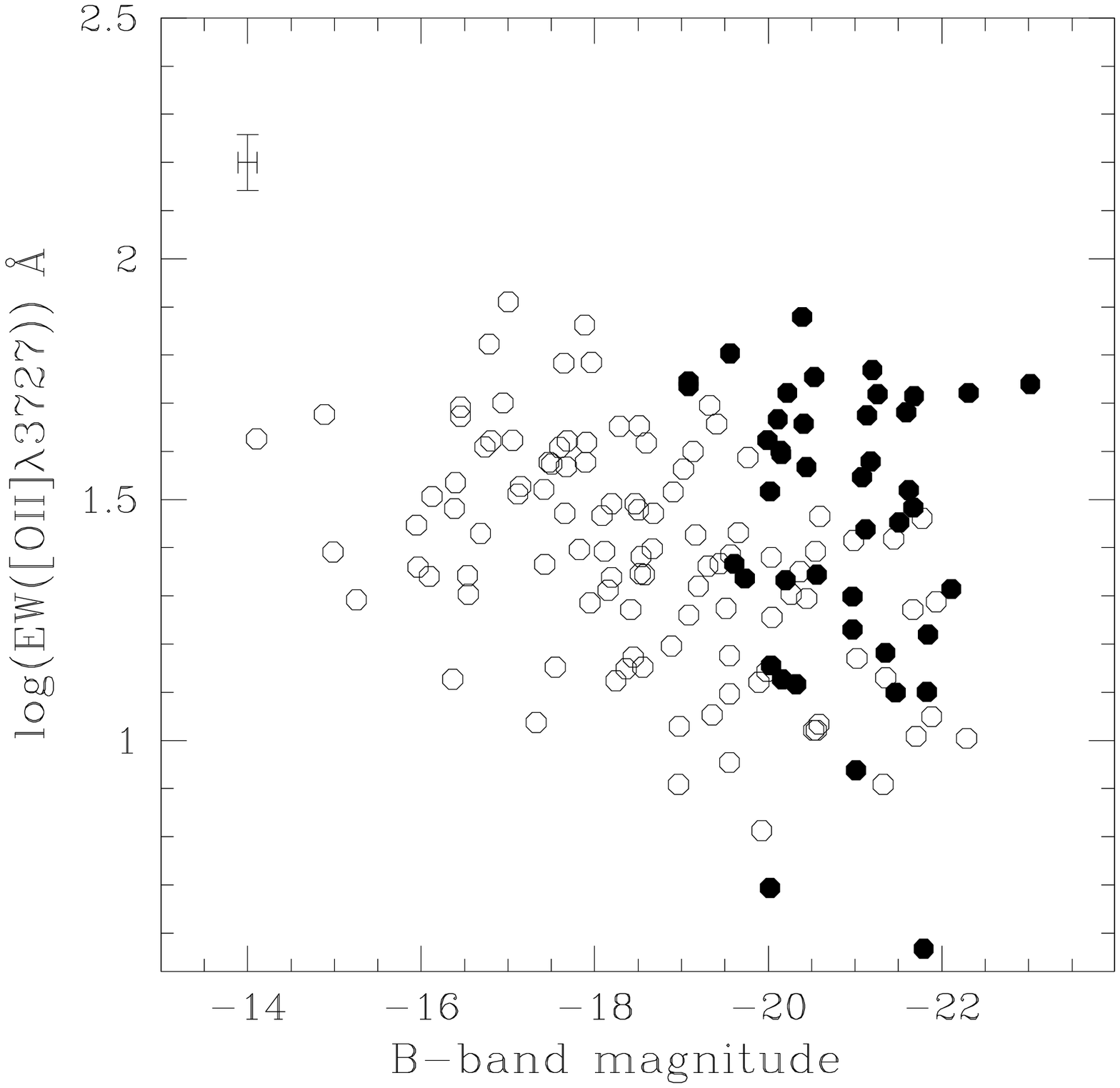}
\includegraphics[clip=,width=0.45\textwidth]{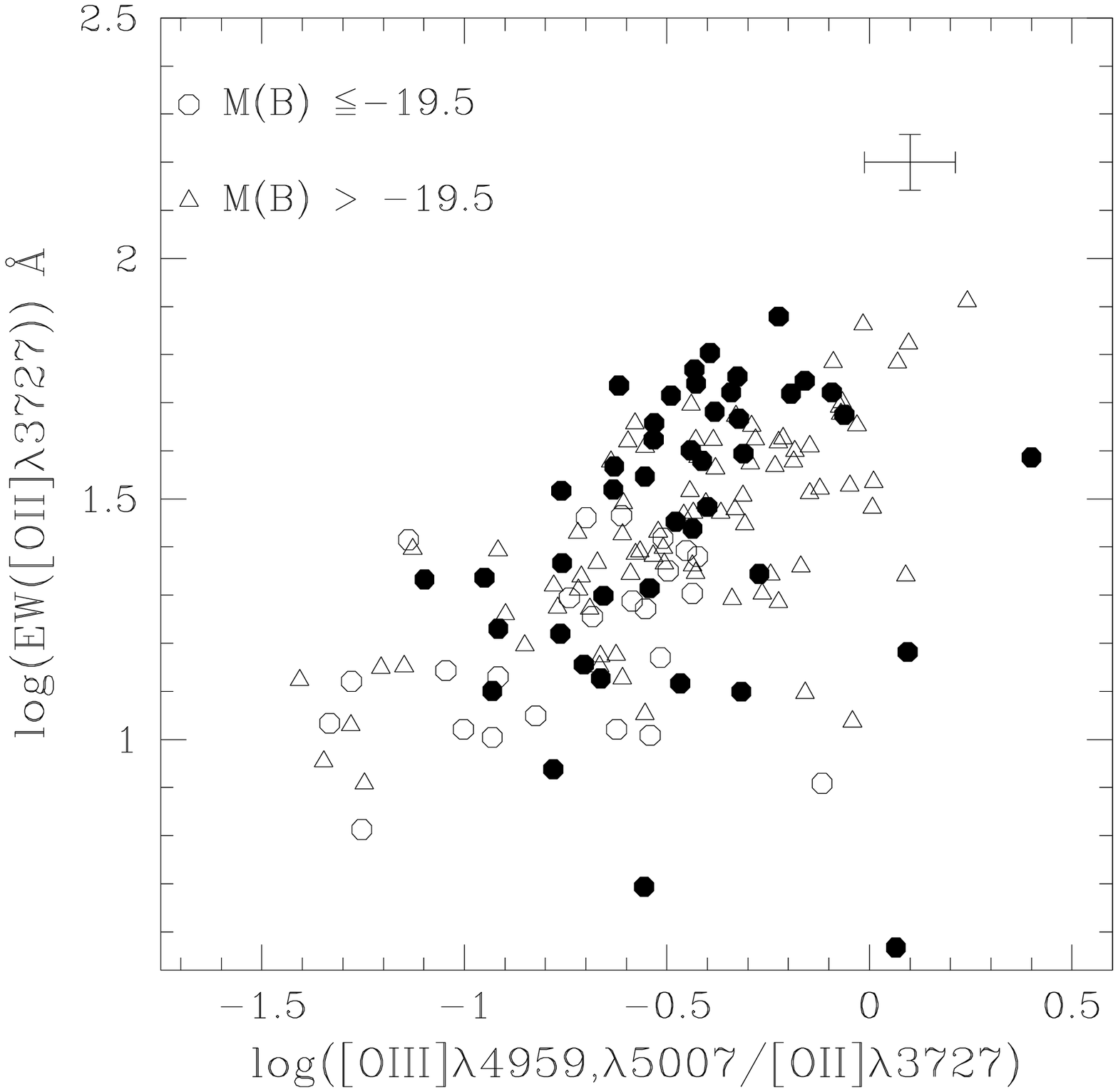}
\caption{{\it Left}: The relationship between rest-frame 
{\oii} emission line equivalent width and B-band absolute 
magnitude for the sample of intermediate redshift 
star-forming galaxies (filled circles), and the NFGS 
sample of local star-forming galaxies (open circles). 
The figure shows clearly that a fraction of intermediate 
redshift galaxies show large {\oii} equivalent widths, 
observed only for faint galaxies at the present epoch. 
{\it Right}: The relationship between rest-frame {\oii} 
emission line equivalent width as a function of the 
excitation-sensitive diagnostic ratio {\oiii}/{\oii}. 
Field star-forming galaxies at intermediate redshifts 
are shown as filled circles, open triangles show faint, 
i.e., $M_{B} > -19.5$, NFGS galaxies, and open circles 
show bright, i.e., $M_{B} > -19.5$, NFGS galaxies. 
The figure shows that a fraction of galaxies in our 
sample occupy the same region as local faint, metal-poor, 
star-forming galaxies.}
\label{ewo2_mb_o32}
\end{figure*}

The left panel of Fig.~\ref{ewo2_mb_o32} shows the 
relationship between galaxy absolute B-band magnitude 
and {\oii} rest-frame emission line equivalent width. 
Galaxies in our sample are shown as filled circles, and 
the local star-forming galaxies in the NFGS sample are 
shown as open circles. 
The NFGS galaxies in this figure shows the well-established 
correlation between galaxy luminosity and emission line 
equivalent width, i.e., the bright end of the galaxy 
luminosity function at the present epoch is dominated 
by galaxies with low emission line equivalent width (e.g., 
Salzer et al. 1989; Kong et al. 2002; Jansen et al. 2000). 
Intermediate redshift star-forming galaxies in our sample 
cover a similar range of {\oii} rest-frame emission line 
equivalent width to that observed locally, but over a much 
narrower luminosity range, i.e., $\sim 2$ mag in comparison 
to the $\sim 7$ mag covered by the NFGS sample. Strikingly, 
a fraction of massive and luminous field galaxies at 
intermediate redshifts show large equivalent widths that 
are seen locally only for faint, i.e., $M_{B} \ga -18$, 
(and metal-poor) galaxies (see also Lilly et al. 2003). 

Locally, faint and metal-poor galaxies that show large 
emission line equivalent widths tend to be highly ionized, 
while bright and metal-rich galaxies are characterized by 
low-ionization parameters (e.g., McCall et al. 1985; 
Stasi\'nska 1990; Mouhcine et al. 2005). The right panel 
of Fig.\,\ref{ewo2_mb_o32} shows the relationship between 
{\oii} emission line rest-frame equivalent width and the 
diagnostic ratio {\oiii}/{\oii}. 
The line ratio {\oiii}/{\oii} is a function of both ionization 
parameter and metallicity (Kewley \& Dopita 2002). 

The {\oiii}/{\oii} ratio has been estimated using emission 
line equivalent widths. Kobulnicky \& Phillips (2003) have 
shown that estimates of this ratio using equivalent widths 
give results similar to using emission line fluxes. 

To illustrate the effect of galaxy luminosity (and hence
metallicity for local galaxies, through their well established 
metallicity-luminosity relation) on the diagnostic ratio 
{\oiii}/{\oii}, we split the local sample of star-forming 
galaxies into faint ($M_{B} > -19.5$) and bright 
($M_{B} \le -19.5$), samples. 
The right panel of Fig.\,\ref{ewo2_mb_o32} shows that local 
star-forming galaxies are distributed along a well-defined 
sequence, interpreted as a metallicity-excitation sequence, 
with metallicity decreasing as the excitation parameter 
increases (e.g., Stasi\'nska 1990). Intermediate redshift 
galaxies in our sample cover a similar range of {\oiii}/{\oii} 
diagnostic ratio as is observed for local star-forming 
galaxies. In addition, star-forming galaxies in our sample 
are distributed similarly to local star-forming galaxies in 
the {\oii} rest-frame equivalent width vs. the diagnostic 
ratio {\oiii}/{\oii} diagram. Interestingly, however, a 
sub-sample of luminous star-forming galaxies at intermediate 
redshifts exhibit {\oiii}/{\oii} ratios that are seen for 
much fainter local galaxies. This suggests that the physical 
parameters of the interstellar star-forming gas within luminous 
intermediate redshift, star-forming galaxies, i.e., their 
metallicity and ionization conditions, might be similar to 
those observed for fainter galaxies that are actively 
forming stars at the present epoch. This behaviour seems 
to extend to higher redshifts. Indeed, Maier et al. (2006) 
have shown that $z \sim 1.4$ luminous star-forming galaxies 
exhibit much higher {\oiii}/{\oii} ratios than galaxies 
with similar luminosities at $0.4 \la z \la 0.9$. Comparable
{\oiii}/{\oii} ratios are found only for much fainter 
star-forming galaxies at this lower redshift range. 
These constraints on the redshift evolution of ionization 
conditions of the interstellar star-forming gas, i.e., galaxies 
with a given {\oiii}/{\oii} ratio are found to be brighter at
earlier times, support the dowssizing scenario for galaxy 
formation (see also Maier et al. 2006; see Sect.\,\ref{disc} 
for more details).

\begin{figure}
\includegraphics[clip=,width=0.45\textwidth]{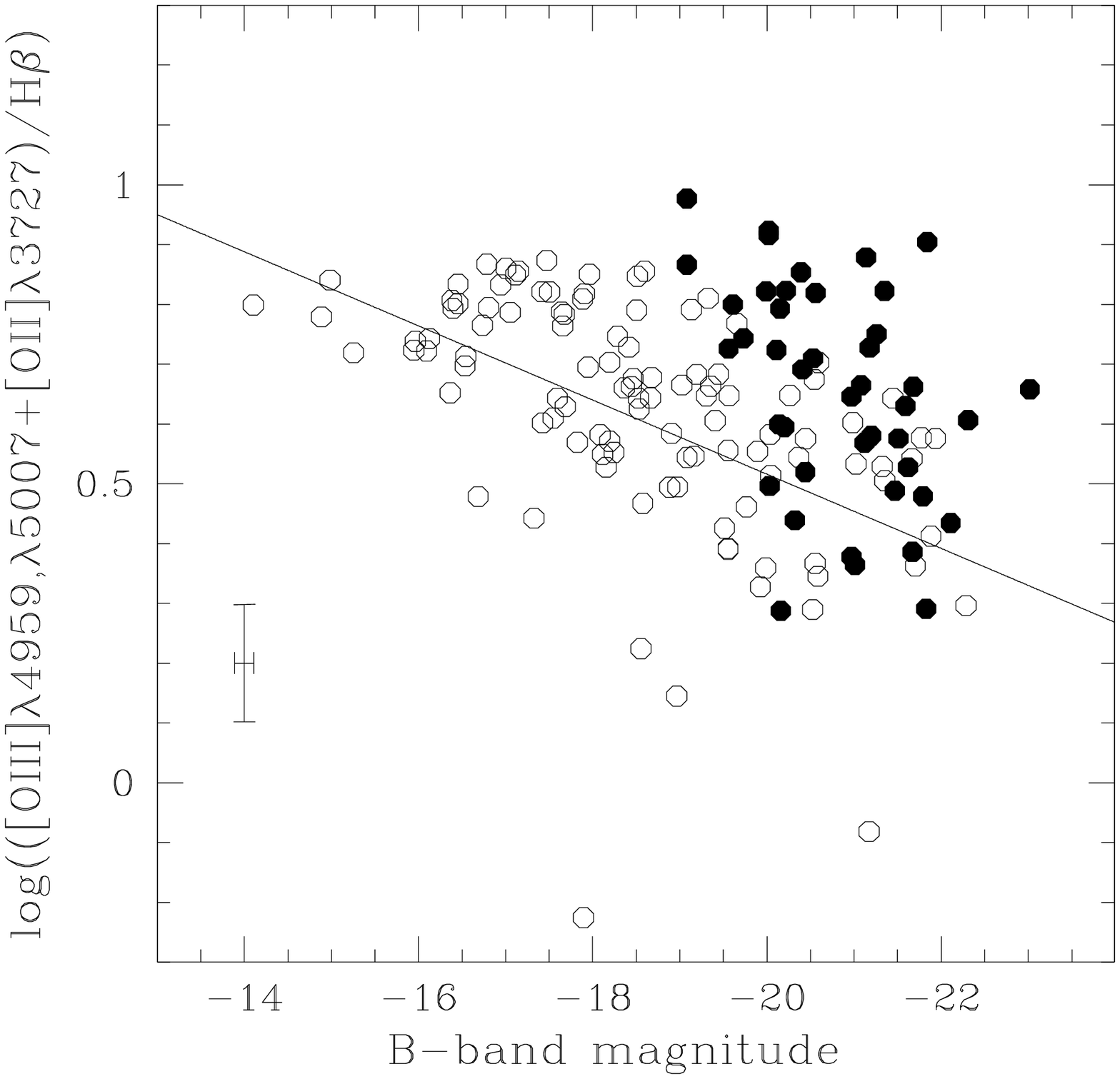}
\caption{The metallicity-sensitive ({\oiii}+{\oii})/{\hbeta} 
ratio versus $M_{B}$ diagram for the sample considered in 
the paper, shown as filled circles, and the local comparison 
sample of star-forming galaxies, shown as open circles. 
The solid line is the fit of Jansen et al. (2001). 
A fraction of luminous (and massive) star-forming galaxies 
at intermediate redshift show large ({\oiii}+{\oii})/{\hbeta} 
ratios indicative of low oxygen abundances. }
\label{r23_mb}
\end{figure}

The ratio of hydrogen recombination lines to collisionally 
excited lines observed in one or more ionization states
of oxygen is used as an abundance-sensitive ratio (e.g.,
Pagel et al. 1979). 
Fig.~\ref{r23_mb} shows the relationship between the most 
commonly used abundance-sensitive ratio 
({\oiii}+{\oii})/{\hbeta}, the so-called {\rp} parameter 
(see Section~\ref{oxy_abund} for more details) and B-band 
absolute magnitude for both local star-forming galaxies 
from the NFGS sample and our sample of intermediate redshift 
field galaxies. 
The solid line is the linear fit to the NFGS galaxy sample 
(Jansen et al. 2001). Again, the measured {\rp} parameter 
for intermediate redshift star-forming galaxies cover a 
similar range to that observed for local star-forming 
galaxies. 

Locally, the abundance-sensitive {\rp} parameter is 
correlated with metallicity, i.e., bright galaxies tend to 
have on average lower {\rp} parameters than faint galaxies. 
However, galaxies in our sample show a large scatter of 
{\rp} parameter at a given galaxy luminosity. Thus, unlike
the local sample of star-forming 
galaxies, the sample of star-forming field galaxies at 
intermediate redshift contains luminous and massive objects 
with large {\rp} parameters, which are found locally only 
at lower luminosities, and are indicative of low oxygen 
abundances.

The observed excitation- and abundance-sensitive diagnostic 
ratios for star-forming galaxies in our sample are similar 
to what is exhibited by local galaxies over much larger 
luminosity and abundance ranges, i.e., from faint/metal-poor 
to luminous/metal-rich star-forming galaxies. 
The observed large variation of the interstellar ionized 
gas properties within luminous field galaxies at intermediate 
redshift suggests that these galaxies do not represent a 
homogeneous population of galaxies in terms of the physical 
properties that shape emission-line galaxy spectra, e.g., 
the ionizing stellar populations and the metal content. 
This indicates a sizeable evolutionary change of the 
properties of massive galaxies between $z \sim 0.5$ 
and the present epoch.


\begin{figure}
\includegraphics[clip=,width=0.45\textwidth]{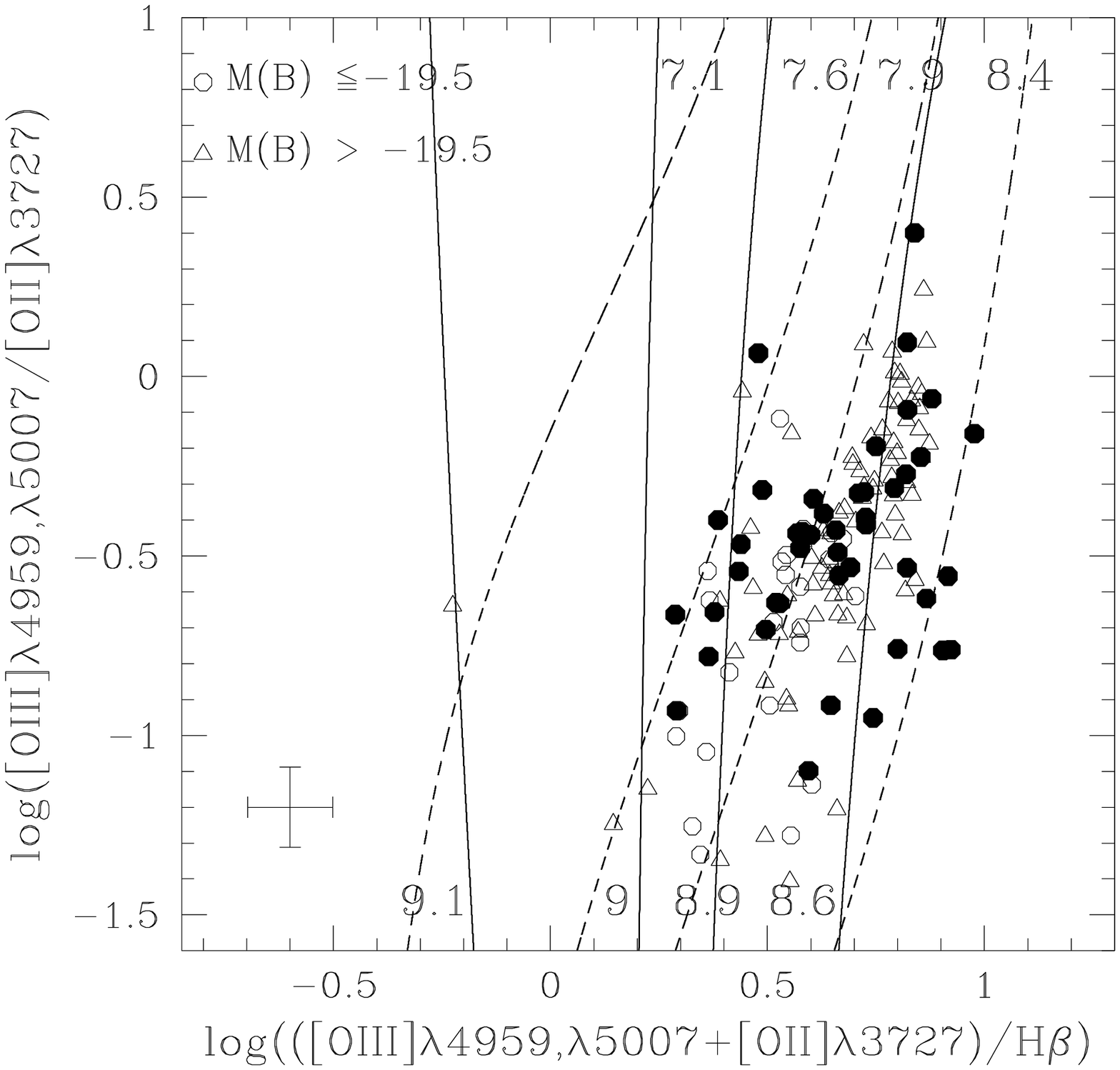}
\caption{The ({\oiii}+{\oii})/{\hbeta} versus {\oiii}/{\oii} 
diagram for the sample considered in this paper. Also shown 
is the McGaugh (1991) calibration of tracks with constant 
metallicity. Models with oxygen abundances lower than 
${\doh}\la\,8.4$ are shown as dashed lines, those with higher 
oxygen abundances are shown by continuous lines. Symbols are
the same as in the right panel of Fig.\ \ref{ewo2_mb_o32}.}
\label{o32_r23}
\end{figure}

\begin{figure*}
\includegraphics[clip=,width=0.45\textwidth]{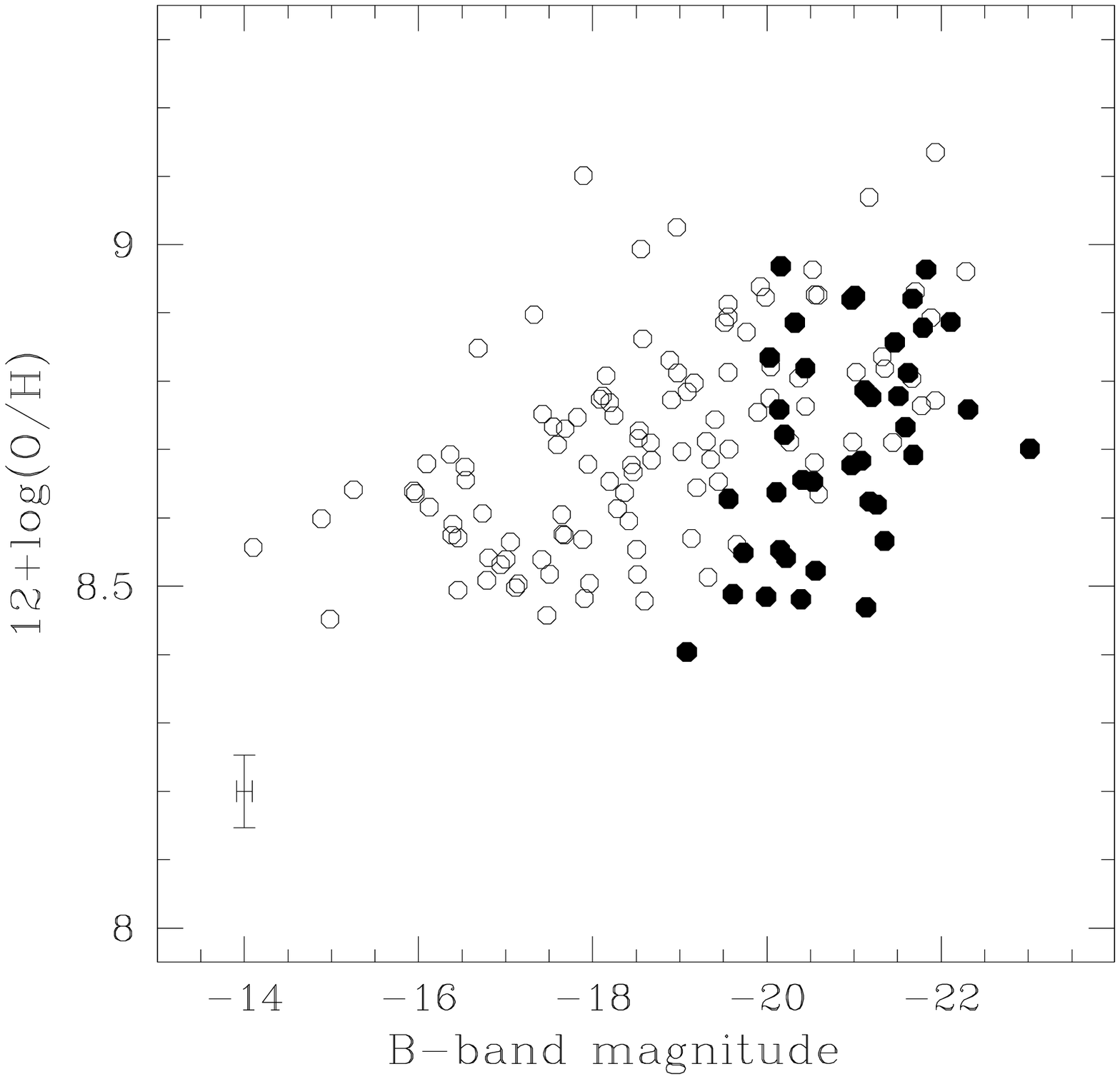}
\includegraphics[clip=,width=0.45\textwidth]{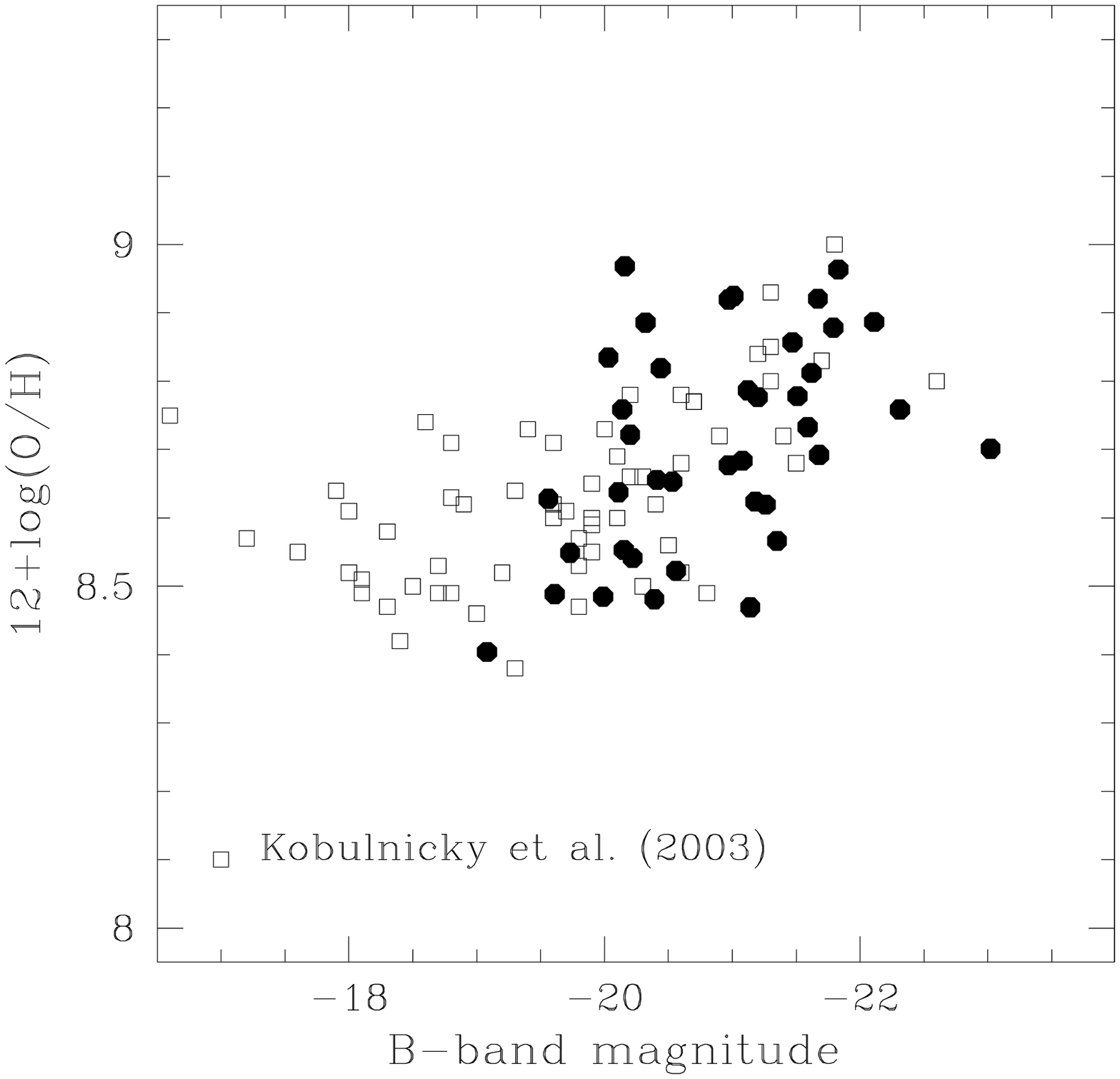}
\includegraphics[clip=,width=0.45\textwidth]{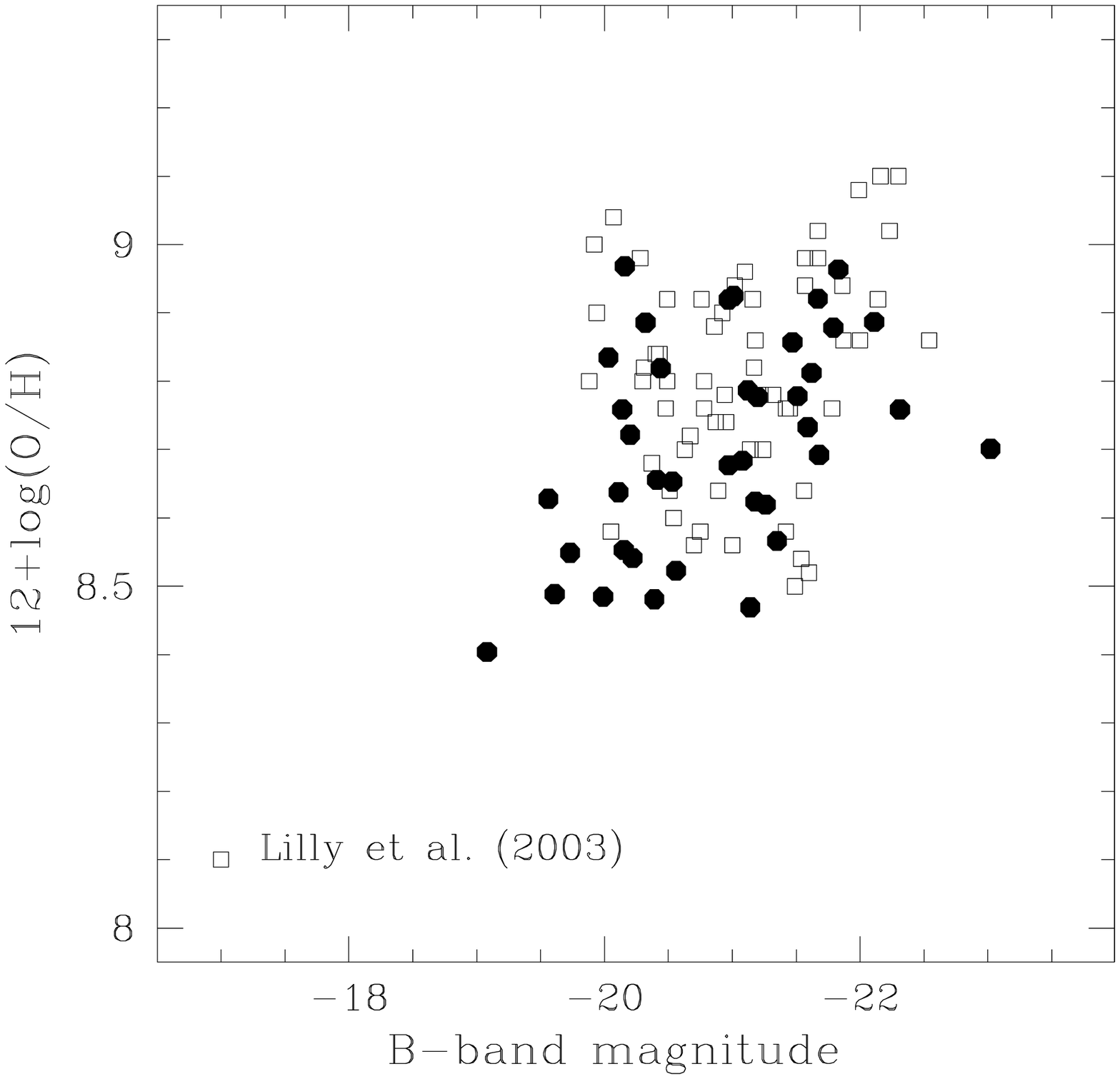}
\includegraphics[clip=,width=0.45\textwidth]{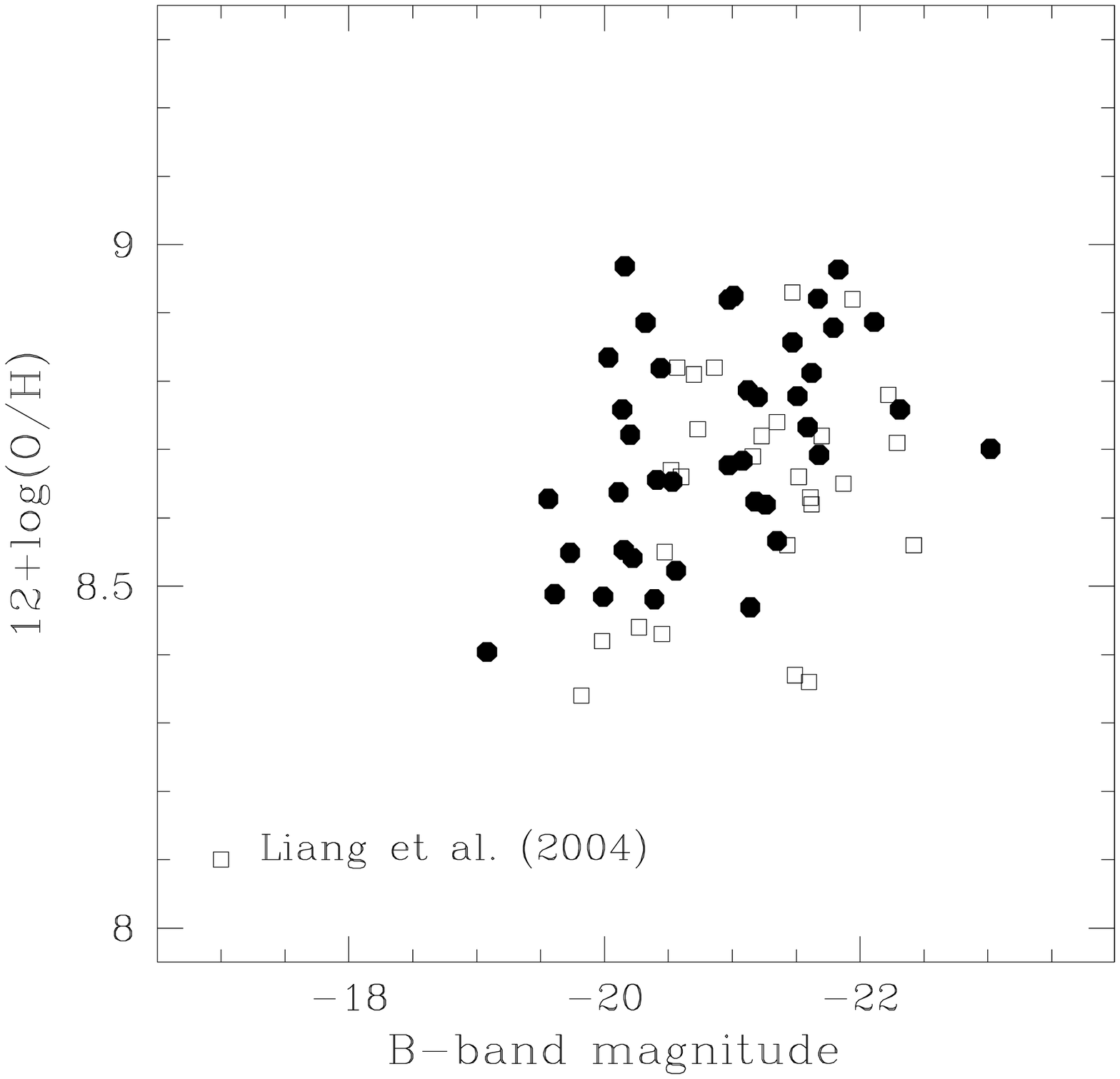}
\caption{Luminosity--metallicity relation for our sample of 
intermediate redshift star-forming galaxies (filled circles), 
compared with several other samples, shown as open symbols.
The comparison samples are: (upper-left) local star-forming
galaxies from the NFGS sample; (upper-right) intermediate
redshift galaxies, $0.4<z<0.82$, from Kobulnicky et al. (2003); 
(lower-left) the galaxy sample of Lilly et al. (2003), with 
$0.47<z<0.92$; and (lower-right) luminous infrared galaxies 
with $0.4<z<1.16$ from Liang et al. (2004). Oxygen abundances 
have been estimated assuming that all star-forming galaxies 
are on the upper-branch of the relationship between {\rp} 
parameter and {\doh} (see text for more details).}
\label{lz_interz}
\end{figure*}

\begin{figure*}
\includegraphics[clip=,width=0.45\textwidth]{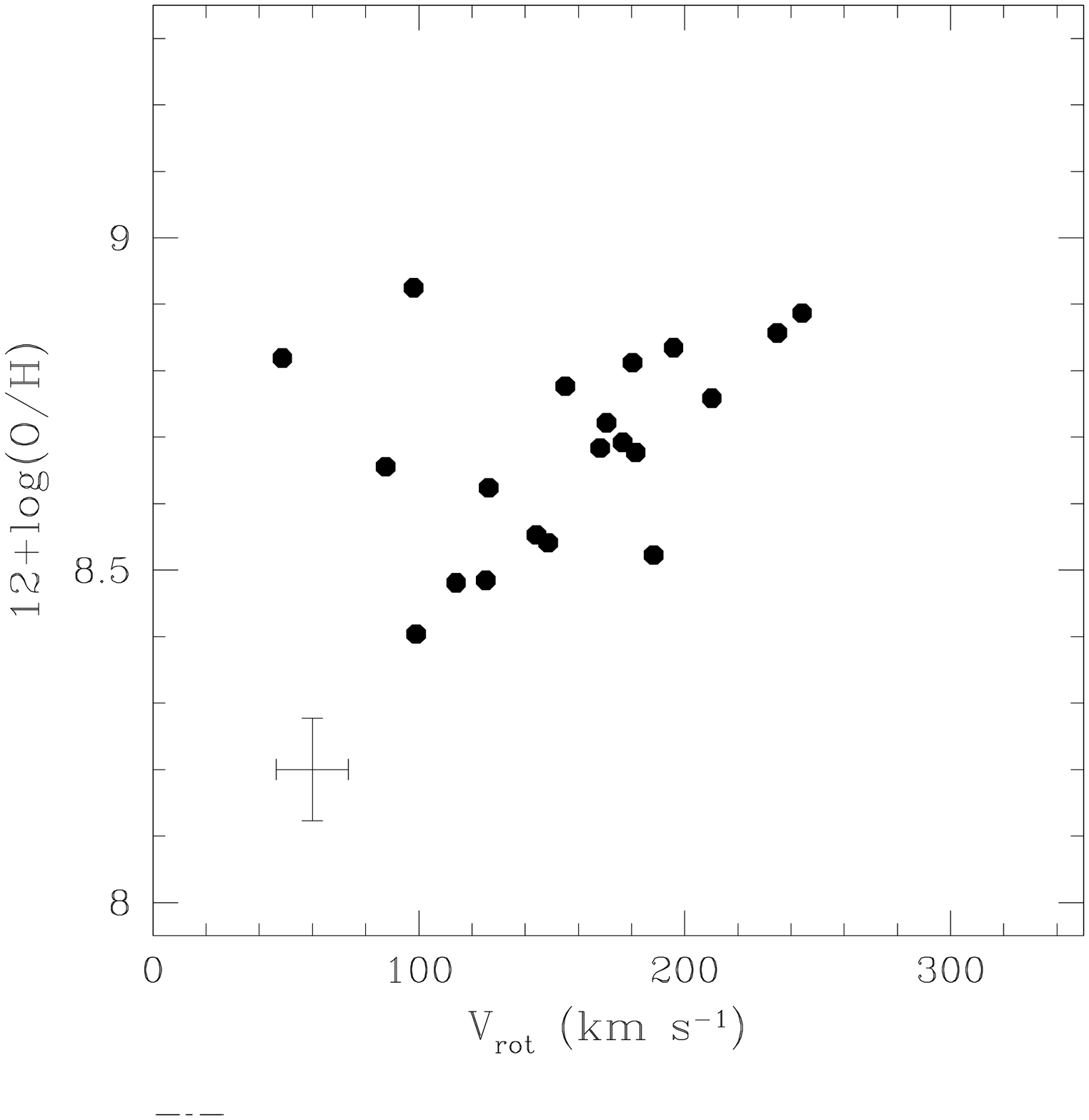}
\includegraphics[clip=,width=0.45\textwidth]{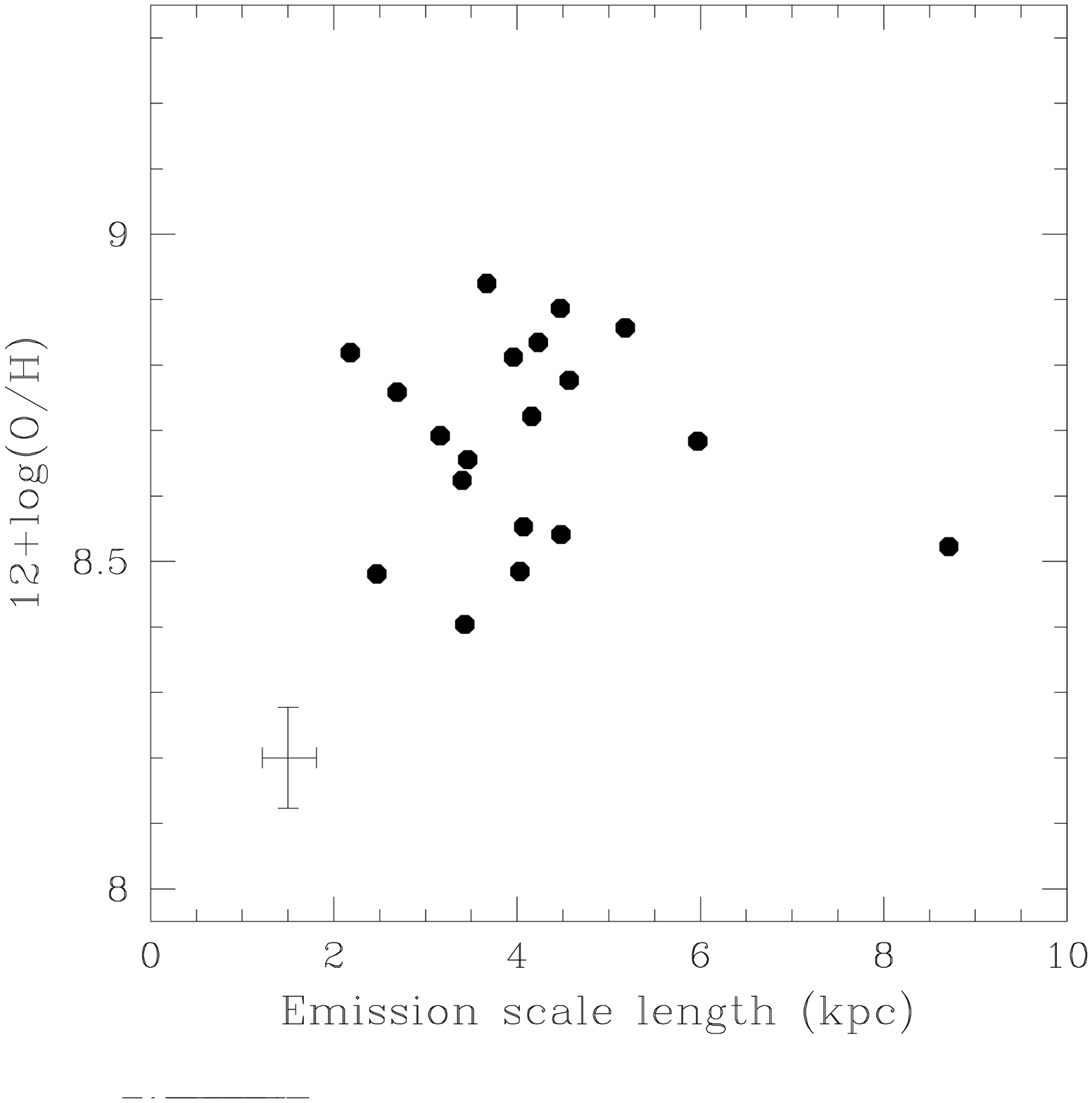}
\caption{Relationship between oxygen abundance in terms of 
{\doh}, estimated using {\rp} parameter and assuming that 
all star-forming galaxies lie on the upper branch of {\rp} 
vs. {\doh} calibration, and maximum rotation velocity 
(left panel), and galaxy size (right panel). No convincing 
correlations are seen.}
\label{oh_vrot_size}
\end{figure*}

\begin{figure*}
\includegraphics[clip=,width=0.45\textwidth]{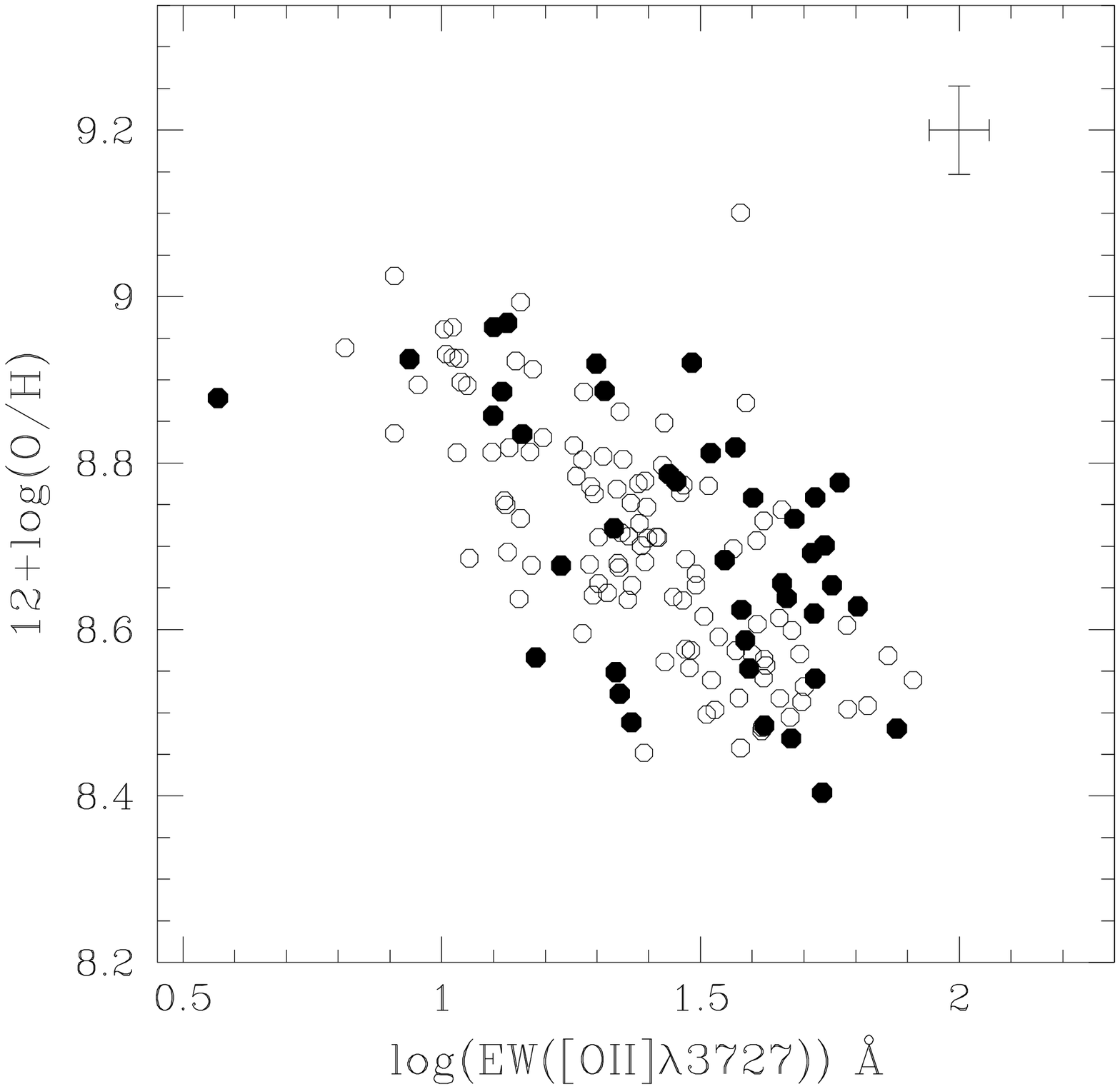}
\includegraphics[clip=,width=0.45\textwidth]{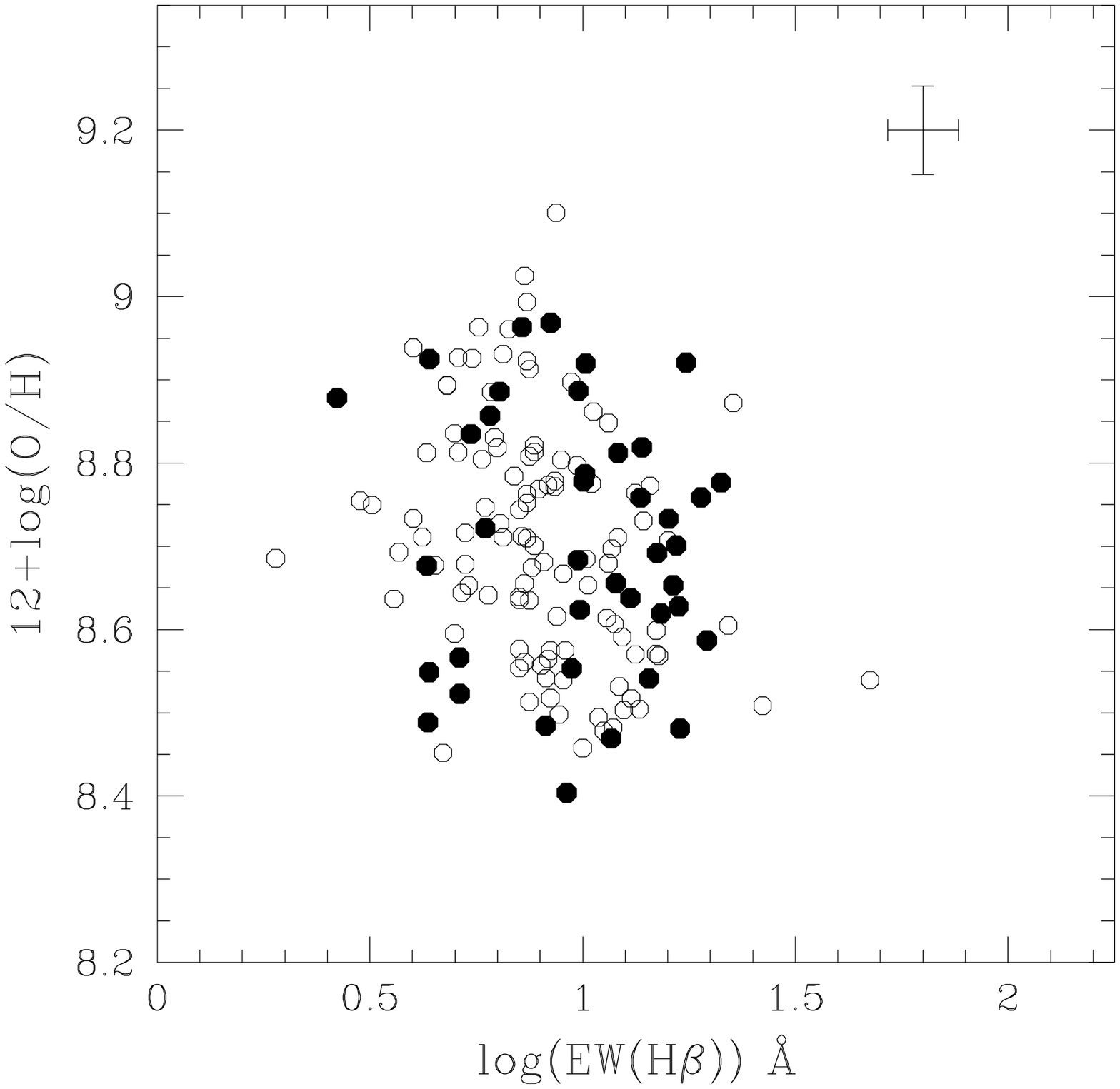}
\caption{Relationship between oxygen abundance in terms of 
{\doh}, estimated using {\rp} parameter and assuming that 
all star-forming galaxies lie on the upper branch of {\rp} 
vs. {\doh} calibration, and rest-frame {\oii} (left panel) 
and {\hbeta} (right panel) emission line equivalent widths. 
Intermediate redshift star-forming galaxies are shown as 
filled circles, and the NFGS sample of local star-forming 
galaxies is marked by open circles.}
\label{oh_ew}
\end{figure*}

\begin{figure}
\includegraphics[clip=,width=0.45\textwidth]{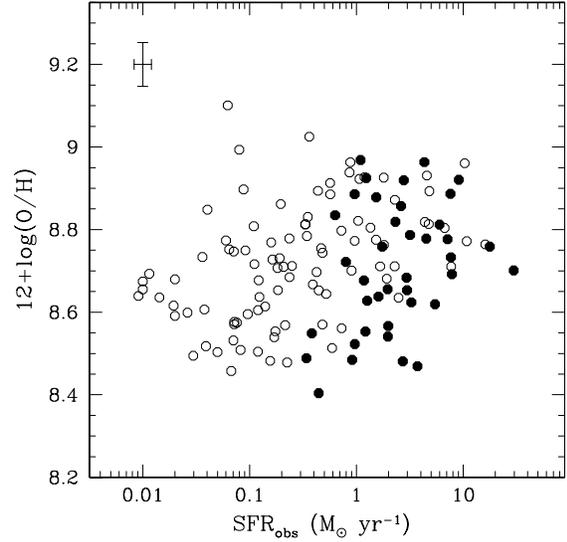}
\caption{Relationship between oxygen abundance in terms of 
{\doh} and observed star formation rate for the sample of 
intermediate redshift star-forming galaxies (filled circles), 
and the NFGS sample of local star-forming galaxies (open 
circles). The current star formation rate has been estimated 
for our galaxy sample using {\hbeta} emission line luminosity 
corrected for stellar absorption, but not for internal dust 
obscuration, and using extinction-uncorrected {\halpha} 
luminosity for the NFGS sample.} 
\label{oh_sfr}
\end{figure}

\begin{figure}
\includegraphics[clip=,width=0.45\textwidth]{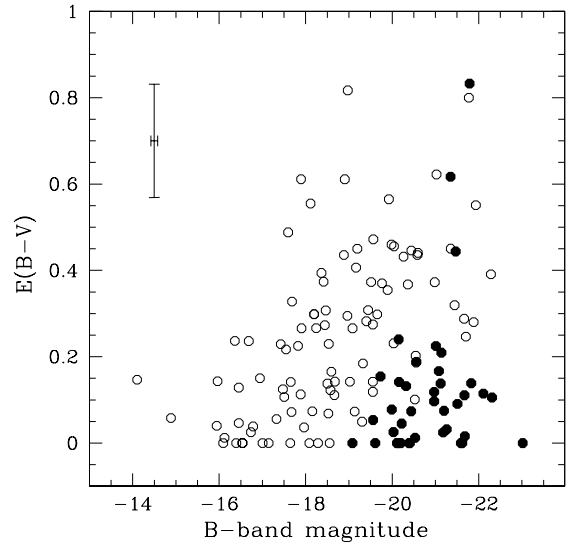}
\caption{Relationship between colour excess and B-band 
magnitude for the sample of intermediate redshift field
galaxies (filled circles), and the NFGS sample of local 
star-forming galaxies (open circles). No correlation is 
observed for intermediate redshift star-forming galaxies.}
\label{ebmv_mb}
\end{figure}

\begin{figure}
\includegraphics[clip=,width=0.45\textwidth]{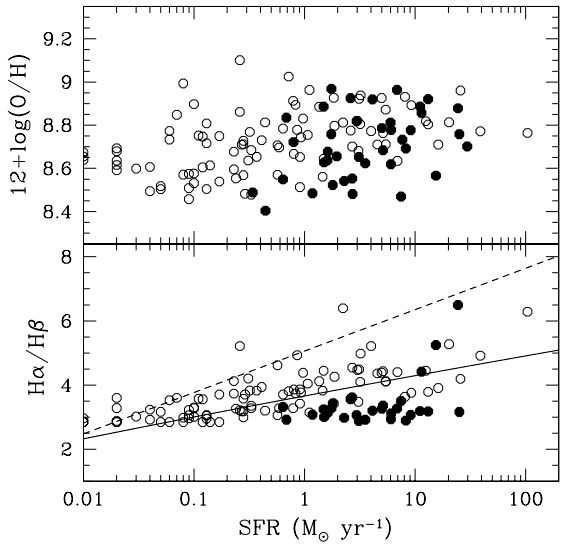}
\caption{Upper panel: the relationship between oxygen 
abundance and extinction-corrected star formation rate for 
the sample of intermediate redshift star-forming galaxies 
(filled circles), and the local sample of star-forming 
galaxies from the NFGS sample (open circles). Lower panel: 
the relationship between Balmer decrement and 
extinction-corrected star formation rate. Symbols are the 
same as for the upper panel. The solid line represents the 
relationship derived by Hopkins et al. (2001) for a sample 
of local star-forming galaxies, while the dashed line 
indicates the relationship for a sample of radio-selected 
galaxies from Afonso et al. (2003). }
\label{hahb_sfr_oh}
\end{figure}

\begin{figure}
\includegraphics[clip=,width=0.45\textwidth]{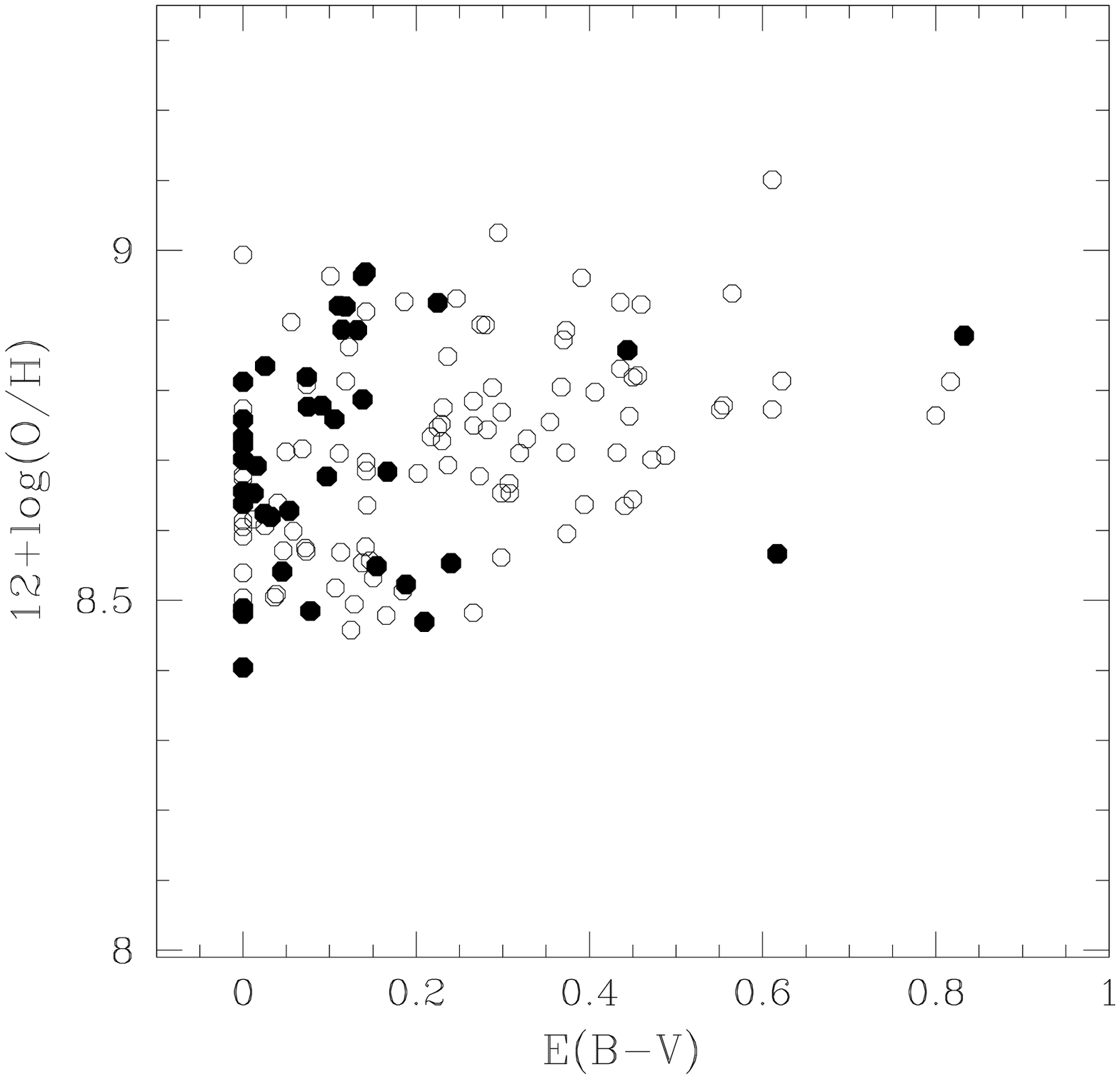}
\caption{Relationship between gas phase oxygen abundance 
and colour excess for the sample of intermediate redshift 
star-forming galaxies (filled circles), and the local sample 
of star-forming galaxies from the NFGS sample (open circles).}
\label{oh_ebmv}
\end{figure}


\section{Properties of luminous, massive intermediate 
redshift galaxies}
\label{prop}

\subsection{Gas phase oxygen abundances}
\label{oxy_abund}

The most reliable method of deriving the gas-phase oxygen 
abundance employs an estimate of the electronic 
temperature and density of the ionized gas (Osterbrock 
1989). An accurate determination of these parameters 
requires reliable measurements of temperature-sensitive 
auroral lines, usually the {\oiiic} emission line. 
Unfortunately, the observed spectra do not have the 
required signal-to-noise to correctly measure this line. 

The absence of directly detected auroral lines, 
particularly in the case of metal-rich galaxies where 
these lines are too weak to be observed, requires the 
development of alternative methods based on strong 
emission lines. The most widely used method is based
on measurements of {\oii}, {\oiii}, and {\hbeta}.
These lines contain enough information to get an accurate 
estimate of the oxygen abundance (McGaugh 1991). This is 
done through the parameter {\rp} introduced initially by 
Pagel et al. (1979), and defined as follows:
\begin{equation}
R_{23}=\frac{{\oiii}+{\oii}}{{\hbeta}}
\end{equation}
                                                        
Extensive studies have been dedicated to calibrating the 
relation between {\rp} and oxygen abundance (e.g., McCall 
et al. 1985; Pilyugin 2001), and now strong-line ratios 
can reliably indicate the oxygen abundance to within the 
accuracy of the model calibrations, approximately 
$\pm 0.15$ dex. The {\rp} parameter is both abundance- 
and ionization-sensitive (e.g., Kewley \& Dopita 2002). 
The correction of this dependence of {\rp} parameter on 
ionization is usually done by using the ionization-sensitive 
diagnostic ratio {\oo}\,=\,{\oiii}/{\oii} (e.g., McGaugh 
1991; Kewley \& Dopita 2002). Generally, the {\oiiib} 
emission line is of a low signal-to-noise. We thus use 
the theoretical ratio {\oiiis}/{\oiiib}\,=\,3 to estimate 
the {\oiiib} equivalent width (Osterbrock 1989).
 
The galaxy spectra analysed here are not flux-calibrated, 
so we use emission line equivalent widths to estimate 
both the {\rp} and {\oo} parameters. We follow the prescription
of Kobulnicky \& Phillips (2003) for estimating the gas phase 
oxygen abundances using strong emission line equivalent widths.
Further analysis using flux-calibrated spectra of both local 
and intermediate redshift galaxies have shown that, for a large 
range of nebular reddening, {\rp}-based oxygen abundances and 
ionization-sensitive ratios based on emission line equivalent 
widths offer comparable precision to estimates using 
reddening-corrected fluxes (see Moustakas \& Kennicutt 2006 
and Lamareille et al. 2006 for more details). This suggests 
that {\rp} and {\oo} parameters are not significantly affected 
by the differential reddening between the stars and the emitting 
gas.

The strong line method has been suspected to involve 
systematic errors, being biased toward higher values 
of oxygen abundances than the method based on 
temperature-sensitive auroral lines, especially at the 
high-metallicity end (Kennicutt et al. 2003). The debate 
over whether the discrepancies in the abundance scales 
are due to systematic biases in the electronic 
temperature-based scale, or in the {H{\sc ii}} region 
models used to calibrate different strong-line vs. 
abundance relations, particularly in the high-metallicity 
regime, is not yet settled. Note that the oxygen abundance 
estimated using different calibrations available in the 
literature might differ by factors of up to $\sim 4$ (Ellison 
\& Kewley 2005). However, as our main interest is to study 
the relative change in oxygen abundance between star-forming 
galaxies at the present epoch and those at intermediate 
redshifts, the exact choice of the {\doh} versus {\rp} 
calibration is not a critical issue. Here we determine 
oxygen abundance using the calibration of McGaugh (1991), 
as found in Kobulnicky, Kennicutt \& Pizagno (1999). 

Fig.~\ref{o32_r23} shows the relationship between 
{\rp} parameter and {\oiii}/{\oii} ratio for star-forming 
galaxies in both the local NFGS sample, and at intermediate 
redshifts. To show the effect of galaxy luminosity, we 
split the local sample of star-forming galaxies into faint
($M_{B} > -19.5$) and bright ($M_{B} \le -19.5$) subsamples.
The McGaugh (1991) calibration of the 
relationship between {\rp} and {\oiii}/{\oii} is overplotted.
Different lines show this relationship for different oxygen 
abundances, in terms of {\doh}. 
The figure shows that star-forming galaxies in both the local 
and intermediate redshift samples are distributed similarly 
in the {\rp} parameter versus {\oiii}/{\oii} ratio diagram.
As shown in the previous section, a fraction of luminous 
intermediate redshift field galaxies are located in a similar 
area as faint local star-forming galaxies with 
intermediate ionization-sensitive {\oiii}/{\oii} ratios and 
{\rp} parameter. Similar to what is found by Lilly et al. 
(2003) for a {\hbeta}-selected galaxy sample and Liang et al. 
(2004) for a sample of luminous infrared galaxies, none of 
the intermediate redshift galaxies in our sample show a 
{\oiii}/{\oii} ratio significantly larger than unity, as 
observed for some extragalactic {H{\sc ii}} regions in 
nearby galaxies (e.g., van Zee et al. 1998) and a few 
$z \ga 2$ luminous galaxies (Kobulnicky \& Koo 2000, 
Pettini et al. 2001, Lemoine-Busserolle et al. 2003).
The sample of intermediate redshift star-forming galaxies 
studied by Maier et al. (2004) does contain a few galaxies 
with {\oiii}/{\oii} larger than unity.
However, all these intermediate redshift highly ionized 
galaxies are 2--3 magnitudes fainter than the galaxies 
in our sample. 

A well-known complication of the use of {\rp} to 
estimate the oxygen abundance is that the dependency 
on metallicity of this parameter is degenerate. 
Indeed, at a fixed value of {\rp} two different 
values of metallicity are possible: at the same oxygen
abundance, different ionization parameters lead to 
different values of {\rp} (McCall et al. 1985).
{\rp} increases with oxygen abundance in the low-metallicity 
regime (${\doh} \la 8.2$; in this scale the solar value
is 8.69; Allende Prieto et al. 2001), while for metal-rich 
objects ($\doh \ga 8.4$) it decreases with O/H, reflecting 
the efficiency of oxygen cooling over abundance in these 
objects. In the ``intermediate'' metallicity region 
($8.2 \la$ {\doh} $\la 8.4$), galaxies may have a large 
range of metallicities for a narrow range of {\rp}. 
The uncertainties in this metallicity domain, i.e. whether 
an object with a given {\rp} parameter lies on the 
metal-rich branch or on the metal-poor branch of the 
{\doh} vs. {\rp} relation, dominate the uncertainties 
related to model calibrations.

A variety of abundance indicators have been used to break 
the degeneracy, e.g. {\nii}/{\oiiib} (Alloin et al 1979), 
{\nii}/{\oii} (McGaugh 1994), {\nii}/{\halpha} (van Zee 
et al. 1998, Denicol\'o et al. 2002).
Unfortunately, the data available is not enough to break 
the {\rp} degeneracy, i.e., none cover the {\halpha} and 
{\nii} spectral range. However, the majority of intermediate 
redshift star-forming galaxies with $M_{B} \la -20$,
for which the {\rp}--metallicity degeneracy has been broken, 
are found to lie on the metal-rich branch of the {\doh} 
versus {\rp} calibration (Kobulnicky et al. 2003; 
Lilly et al. 2003; Kobulnicky \& Kewley 2004, 
Maier et al. 2005). While the exact location of galaxies 
in our sample on the {\doh} versus {\rp} calibration 
cannot be determined reliably until near-infrared 
spectroscopy is available for our sample galaxies, we make 
the reasonable assumption that the luminous star-forming 
galaxies in our sample lie on the upper metallicity branch 
of the {\doh} versus {\rp} calibration. The expected oxygen 
abundances of intermediate redshift massive galaxies in 
the case where they lie on the lower metallicity branch 
are extremely low, suggesting improbably large evolutionary 
changes in the metallicities of massive galaxies between 
$z \sim 0.5$--$1.0$ and the present epoch (see Lilly et al. 
2003; Ellison et al. 2005).

For four galaxies in our sample, the oxygen abundance 
estimated using the low-branch of metallicity calibration 
was larger than the one derived using the upper-branch 
of the metallicity calibration. These galaxies have been 
excluded from the sample. This may happen when the {\rp} 
is larger than the maximum value predicted by the model 
grid used to calibrated the relation between {\doh} and 
{\rp}. The eighth column of Table~\ref{gal_prop} lists 
the estimated oxygen abundances.

Fig.~\ref{lz_interz} shows the relationship between galaxy 
B-band absolute magnitude and oxygen abundance in terms of 
{\doh} for our sample of intermediate redshift star-forming 
galaxies compared with several other samples. The comparison 
samples are the local relation defined by star-forming 
galaxies from the NFGS sample, and three intermediate 
redshift samples for which the gas-phase oxygen abundances
were estimated using the strong emission line method, similar 
to our procedure. These are: the {\hbeta}-selected sample of 
Lilly et al. (2003), a sample of star-forming galaxies drawn 
from the Deep Extragalactic Evolutionary Probe Survey of Groth 
Strip galaxies (Kobulnicky et al. 2003), and the sample of 
intermediate redshift luminous infrared galaxies of 
Liang et al. (2004). All star-forming galaxies in the
NFGS sample are assumed to lie on the upper metallicity 
branch of the {\doh} vs. {\rp} calibration, as was 
assumed for the intermediate redshift galaxy samples. 
As expected from the comparison of different diagnostic 
ratios observed for our sample galaxies with what is 
seen for local star-forming galaxies with known oxygen 
abundances (see section \ref{diag_sect}), the estimated 
oxygen abundances for galaxies in our sample cover a 
large range, i.e., $8.4 \la${\doh}$\la 9$, similar 
to that seen for star-forming galaxies in the NFGS sample, 
with a mean of 8.7.

The upper-left panel of Fig.~\ref{lz_interz} shows 
the well-established correlation between luminosity and 
oxygen abundance for galaxies in the local universe (e.g., 
Melbourne \& Salzer 2002; Lamareille et al. 2004;
Tremonti et al. 2004), with a large scatter, however, 
which could be related to the variety of star formation 
histories in these galaxies (see e.g., Mouhcine \& Contini 
2002). Unlike local galaxies, a clear correlation between 
galaxy luminosity and oxygen abundance is not observed for 
our sample of star-forming field galaxies at intermediate 
redshifts. Rather, a broad range of oxygen abundances of 
the interstellar star-forming gas at any given galaxy 
luminosity is observed. Our sample contains luminous 
galaxies that exhibit oxygen abundances similar to those 
seen for bright galaxies at the present epoch, i.e., 
${\doh} \sim 8.8$--$9$, suggesting that the chemical 
properties of this population of intermediate redshift 
galaxies will not evolve significantly to the present 
epoch. This is in agreement with recent studies of the 
chemistry of luminous star-forming galaxies at intermediate 
redshifts (e.g., Kobulnicky \& Kewley 2004).

However, a subsample of luminous intermediate redshift 
star-forming galaxies show oxygen abundances lower 
than what is observed for local galaxies with similar 
luminosities, i.e., ${\doh} \sim 8.4$--$8.6$, in 
agreement with the recent finding of Maier et al. (2005). 
Given their rotation velocities and physical sizes, the 
population of metal-poor galaxies in our sample is likely 
to evolve into massive metal-rich galaxies in the local 
universe. 
Their evolution in the luminosity vs. metallicity diagram 
will be vertical rather than horizontal. Thus, these 
galaxies are on the process of building up their metal 
content. This is in agreement with the suggested scenario 
of Lilly et al. (2003), who argue based on galaxy 
properties, i.e., optical and near-infrared photometry, 
morphological properties and oxygen abundances, that 
bright intermediate redshift galaxies with intermediate 
oxygen abundances are likely to evolve into metal-rich 
disk galaxies at the present epoch by a progessive 
enrichment of their metal content, rather than fade into 
local dwarf galaxies. The appearance of a population of 
luminous and metal-poor galaxies at intermediate redshifts 
should affect the luminosity-metallicity relation at this 
redshift range, in the sense that galaxies of a given 
luminosity become, on average, more metal-poor at higher 
redshift (see e.g., Kobulnicky \& Kewley 2004).

The upper-right panel of Fig.~\ref{lz_interz} compares
the luminosity--metallicity relation of our sample of 
intermediate redshift galaxies with the sample of 
galaxies from Kobulnicky et al. (2003), with $0.26<z<0.82$. 
Note that Kobulnicky et al. (2003) used emission line 
equivalent widths rather than fluxes to estimate oxygen
abundances, similar to our procedure. 
The oxygen abundances of both samples cover similar ranges, 
however the sample of Kobulnicky et al. (2003) extends 
to fainter luminosities. Unlike our sample galaxies, the 
intermediate redshift galaxy sample from Kobulnicky et al. 
(2003) shows a trend between galaxy luminosity and gas 
phase oxygen abundance.
Galaxies in our intermediate redshift sample cover a wider 
range of oxgen abundance at a given luminosity. Overall the 
Kobulnicky et al. (2003) galaxies tend to be fainter at a 
given oxygen abundance.  However, this difference simply 
reflects that our sample is limited to brighter galaxies 
than the rest-frame B-band selected sample from which the 
Kobulnicky et al. (2003) sample has been drawn (see also 
Liang et al. 2004).

The lower-left panel of Fig.~\ref{lz_interz} compares 
the luminosity--metallicity relation of our intermediate 
redshift star-forming galaxies with a sample of
Canada-France Redshift Survey galaxies with $0.47<z<0.92$
from Lilly et al. (2003). Both samples show similar 
luminosity--metallicity relations.

Liang et al. (2004) investigated the luminosity--metallicity 
relation for a sample of intermediate redshift, luminous
infrared (${\rm 15 \mu m}$-selected) galaxies with 
$0.4<z<1.16$. The comparison between the 
luminosity--metallicity relation for our sample galaxies 
and the sample of luminous infrared galaxies of Liang et al. 
(2004) is shown in the lower-right panel of Fig.~\ref{lz_interz}. 
Interestingly, both galaxy samples show similar relations, 
despite the fact that optically selected galaxy samples
and mid-infrared selected galaxy samples tend to select 
different population of galaxies, i.e., luminous infrared
galaxies in the Liang et al. (2004) sample tend to show on 
average larger internal dust obscuration and intrinsic 
star formation rates than galaxies in our sample (see
Section~\ref{sfr_ebmv} for a discussion of the distribution
of the star formation rate and colour excess for the 
galaxies in our sample). The similarity of the observed
luminosity--metallicity relations might be due to that 
luminous infrared galaxies at intermediate redshifts 
showing similar rest-frame optical and spectral properties 
to normal starburst galaxies, and suggests the differences
in the selection procedure do not lead to a significant
bias against a given class of galaxies (Hammer et al. 
2005).

If the assembly of galaxy stellar mass, and subsequently
its metal content, is related to the size of the dark 
matter halo where a galaxy sits, the degree of the 
interstellar gas chemical evolution is expected to 
correlate with maximum rotation velocity of the halo.
Fig.~\ref{oh_vrot_size} shows the relations between 
oxygen abundance in terms of {\doh} and rotation velocity 
and emission length scale, shown in the left and right 
panels, respectively, for our sample of star-forming 
field galaxies. No convincing correlations are seen 
between gas phase oxygen abundance and rotation velocity 
or emission length scale. For a given rotation velocity 
or galaxy size, intermediate redshift galaxies 
exhibit a wide range of gas phase oxygen abundances. 
This is consistent with the lack of a correlation 
between gas phase oxygen abundance of luminous 
intermediate redshift star-forming galaxies and their 
I-band half-light radius reported by Lilly et al. (2003). 
This contrasts however with the reported positive trend 
between gas phase oxygen abundance and rotation velocity 
for local spiral galaxies (Zaritsky et al. 1994). 
This suggests that at least a fraction of massive and 
large galaxies at intermediate redshifts continue the 
assembly of their chemical content after $z \sim 1$.
Metal-rich and metal-poor galaxies cover similar ranges 
of rotation velocity and physical size, implying that 
both galaxy populations reside within similar halos, 
and that the size of the halo is not the only parameter 
that affects galactic chemical enrichment history.

\subsection{Star formation and dust obscuration}
\label{sfr_ebmv}

The population of luminous field galaxies at intermediate 
redshifts consists of a mixture of galaxies with 
different levels of chemical evolution and interstellar
gas properties. This suggests that intermediate redshift 
luminous galaxies might have a variety of star formation 
histories, and thus could exhibit further differences in
their intrinsic properties.

Fig.~\ref{oh_ew} shows the relationship between oxygen 
abundance in terms of {\doh} and rest-frame equivalent 
width of the {\oii} and {\hbeta} emission lines respectively. 
A clear correlation is visible between gas phase oxygen 
abundance and {\oii} emission line rest-frame equivalent 
width. However, no convincing trend is observed between 
oxygen abundance and {\hbeta} emission line rest-frame 
equivalent width as also seen by Lilly et al. (2003) 
and Kobulnicky et al. (2003). This could be due to an 
unreliable correction of the underlying stellar Balmer 
lines in absorption which might introduce an extra scatter 
to the relation.

Emission line equivalent widths can be considered as 
rough tracers of the current relative star formation, 
in the sense of absolute current star formation rate 
relative to the total integrated star formation rate, 
i.e., the star formation rate per unit luminosity. 
Intermediate redshift and local star-forming galaxies 
are distributed similarly in oxygen abundance vs. 
emission line rest-frame equivalent width diagrams, 
suggesting that massive, metal-poor field galaxies at 
intermediate redshifts might have similar relative 
star formation histories to local faint and metal-poor 
star-forming galaxies. The observed trend might be 
taken to suggest that systems with low metallicities
on average show higher relative star formation rates 
than metal-rich galaxies. The absence of a correlation 
between oxygen abundance and rotation velocity suggests 
that this might be the case independent of the size 
of the galaxy halo.

Fig.~\ref{oh_sfr} shows the relationship between oxygen 
abundance and star formation rate for star-forming 
galaxies in both the local NFGS comparison galaxy sample, 
shown as open circles, and our galaxy sample, shown as 
filled circles. The star formation rates of intermediate 
redshift galaxies are derived using {\hbeta} luminosities, 
estimated using {\hbeta} equivalent widths and B-band 
absolute luminosities as given by equation~6 of Kobulnicky 
\& Kewley (2004). The star formation rates are then 
calculated following the calibration of Kennicutt (1998). 
The estimated star formation rates, in solar masses per 
year, are listed in the 9th column of 
Table~\ref{gal_prop}. {\hbeta} emission line luminosities 
were not corrected for internal reddening, thus the 
estimated star formation rates of intermediate redshift
galaxies must be considered as lower limits. The star 
formation rate of star-forming galaxies in the NFGS 
sample has been estimated using extinction-uncorrected 
{\halpha} luminosity following the calibration of 
Kennicutt (1998). 

Local star-forming galaxies exhibit a correlation
between oxygen abundance of the interstellar gas and 
the extinction-uncorrected star formation rate, in the 
sense of higher oxygen abundances for galaxies with 
higher observed star formation rate, i.e., observed
{\halpha} luminosity.
However, no clear trend is obvious for intermediate 
redshift galaxies: metal-poor and metal-rich luminous 
star-forming galaxies cover a similar range of 
extinction-uncorrected star formation rate. 
Metal-rich luminous star-forming galaxies at intermediate 
redshifts show, at a given oxygen abundance, similar 
extinction-uncorrected star formation rates to their 
local counterparts. This could be the case if metal-rich 
luminous galaxies at intermediate redshift are affected 
by a similar amount of internal dust obscuration as 
their counterparts at the present epoch. Alternatively,
this could be the case if they show, at a given gas 
phase oxygen abundance, higher/lower Balmer intrinsic 
emission line fluxes and lower/higher internal reddening 
than metal-rich star-forming galaxies locally. On the other 
hand, metal-poor luminous galaxies at intermediate redshift 
show on average larger extinction-uncorrected star formation 
rates than local galaxies with similar gas phase oxygen 
abundances. This might be due to that metal-poor galaxies 
at intermediate redshift have higher intrinsic star 
formation rates or lower internal reddening than their local 
counterparts. To distingle between these possibilities, 
internal dust obscuration affecting intermediate redshift 
star-forming galaxies should be estimated.

As mentioned earlier, none of the objects in our sample
have both {\halpha} and {\hbeta} emission line present 
in their covered spectral range, so the dust obscuration 
cannot be derived using the Balmer decrement. 
However, by comparing the energy balance between the 
luminosities of two different star formation indicators
that are free from systematic effects other than dust 
reddening, i.e., that do not depend on metallicity or 
excitation state of the emitting gas, it is possible 
to estimate the amount of internal extinction affecting 
a galaxy's spectrum. 
Using {\hbeta} and {\oii} emission lines as star formation 
rate indicators, the colour excess can be estimated using 
the relation: 
\begin{equation}
{\rm E(B-V) = \frac{2.5}{\kappa({\hbeta})-\kappa(OII)}
 \log{\left(\frac{SFR(OII)_{obs}}{SFR({\hbeta})_{obs}}\right)}}
\end{equation}
where $\kappa(\lambda)$ is the optical interstellar 
extinction curve. We adopt the Milky Way interstellar 
extinction law of Cardelli, Clayton, \& Mathis (1989), 
with $R_V=3.1$. The {\oii} emission line is a widely used 
star formation indicator for galaxies at the redshift range 
where the {\halpha} emission line moves into the near-infrared 
(e.g., Thompson \& Djorgovski 1991; Cowie et al. 1997; 
Hogg et al. 1998; Hippelein et al. 2003). However, published 
calibrations of the star formation rate in terms of {\oii} 
vary by factors of a few (Gallagher et al. 1989;
Kennicutt 1992; Guzm\'an et al. 1997; Rosa-Gonz\`alez 
et al. 2002). The {\oii} emission line shows a complex 
behaviour as a function of galaxy and interstellar gas 
properties. 
Locally, for star-forming galaxies with 
ionization-sensitive {\oiii}/{\oii} ratios similar to what
is observed for the intermediate redshift galaxies studied 
in this paper, and with ${\doh} \ga 8.4$, the variation of
observed {\oii}/{\halpha} flux ratio is regulated by 
extinction and metallicity, with no sizeable sensitivity 
to the excitation state of the ionized gas (Kewley et al. 
2004; Mouhcine et al. 2005). The star formation rate in 
terms of the {\oii} emission line is estimated using the 
calibration: 
\begin{equation}
{\rm SFR(M_{\odot}\,yr^{-1})\,=\, 7.9\times\,10^{-42}
        \,L_{[OII]} (erg\,s^{-1}) \times\,  f(O/H)}
\end{equation}
where ${\rm L_{\rm [OII]}}$ is the observed {\oii} 
luminosity, and ${\rm f(O/H})$ is a correction factor 
introducted by Kewley et al. (2004) to account for the 
effect of metallicity on the variation of {\oii}/{\halpha} 
flux ratio, i.e., 
${\rm f(O/H)} = 1/[-2.29+21.21 \times ({\doh})]$ (see Kewley 
et al. 2004). 

The {\oii} emission line luminosity is estimated from 
the equivalent width ($W(\rmn{[OII]})$) and the rest-frame 
absolute $B$-band magnitude. For this we estimate the 
continuum flux at the position of the [OII] line from 
$M_B$ and the spectral energy distribution which best 
fits the available colours, which was used to determine 
$M_B$ from the observed magnitudes in the first place.  
The absolute continuum flux at the effective wavelength 
of the $B$-band ($\unisim 4450$\AA) was calculated from 
the rest-frame absolute $B$-band magnitudes using the 
conversion given by Fukugita et al. (1995) $F_{\rmn{cont},B}
= 6.19 \cdot 10^{-9} \cdot 10^{-0.4 M_B}$ 
erg s\per{} cm\per[2]{} \AA\per. Using the spectral 
energy distributions of Arag\'on-Salamanca et al. (1993),
which were used in the determination of the VLT $M_B$ 
and are similar in the range of interest to those used 
for the Subaru $M_B$ (Fukugita et al. 1995), the ratios 
of the continuum flux at {\oii} ($F_{\rmn{cont},[OII]}$) 
to that at the $B$-band effective wavelength
($F_{\rmn{cont},B}$) are $0.45$, $0.53$, $0.67$, $0.74$, 
and $0.83$ for types E/S0, Sab, Sbc, Scd, and Sdm 
respectively (Milvang-Jensen 2003). An approximate
conversion was used to convert the types used for the 
Subaru data onto the system used for the VLT data.  
The ratio was interpolated between the listed values for 
intermediate types. The absolute flux in the {\oii} line
is then simply $F_{\rmn{[OII]}} = W(\rmn{[OII]}) 
\cdot F_{\rmn{cont},[OII]}$.

Since a negative colour excess is unphysical, we assume 
that galaxy properties that give rise to ${\rm E(B-V) < 0}$ 
correspond to zero attenuation. The estimated colour excess 
for our intermediate redshift star-forming galaxies are 
listed in the last column of Table~\ref{gal_prop}.

Fig.~\ref{ebmv_mb} shows the relationship between the 
colour excess and galaxy luminosity for star-forming 
galaxies in both the local comparative sample, and the 
intermediate redshift sample. No correlation is found
between galaxy luminosity and nebular extinction for
intermediate redshift galaxies, in agreement with the
finding of Maier et al. (2005). The estimated colour 
excesses for intermediate redshift galaxies are distributed 
over a range similar to that seen across a large sample of 
local star-forming galaxies with $-14.0 \la M_{B} \la -22.0$. 
However, the mean colour excess for intermediate redshift 
galaxies in our sample, i.e., $0.16 \pm 0.03$ when excluding 
galaxies with ${\rm E(B-V) = 0}$ and $0.12 \pm 0.03$ when
all galaxies are included, is significantly lower than the 
observed average colour excess, $E(B-V) \sim 0.3$, for 
local star-forming galaxies (e.g., Nakamura et al. 2004, 
Hopkins et al. 2003, Kewley et al. 2004, Mouhcine et al. 
2005). Star-forming galaxies at intermediate redshifts 
show a larger scatter of colour excess at a given galaxy 
luminosity than observed locally: a fraction of luminous 
star-forming galaxies at intermediate redshifts show amounts 
of dust obscuration similar to what is observed for their 
local counterparts, while another subpopulation of luminous 
galaxies show dust obscuration similar to that observed 
locally for fainter galaxies. 

The comparison of different star formation rate 
indicators for local samples of star-forming galaxies, 
selected at optical or ultra-violet wavelengths, have 
shown a clear trend for increasing dust obscuration 
with intrinsic star formation rate (e.g., Wang 
\& Heckman 1996; Sullivan et al. 2001; Buat et al. 
2002). The lower panel of Fig.\ref{hahb_sfr_oh} shows 
the relationship between the Balmer decrement, estimated
using the derived E(B-V) and the extinction curve of
Cardelli et al. (1989), and the extinction-corrected 
star formation rate for the local comparative sample, 
shown as open circles, and our sample 
of star-forming field galaxies. The solid line shows the 
relationship derived from Hopkins et al. (2001) for a 
sample of local galaxies, while the dotted line indicates 
the relationship for a radio-selected sample of 
star-forming galaxies to $z \approx 0.8$ from Afonso et al. 
(2003). The distribution of local star forming galaxies in 
the NGFS sample agrees nicely with the relationship from 
Hopkins et al. (2001), however they deviate significantly 
from the relationship observed for radio-selected galaxies. 
This could be because radio-selected galaxy samples tend 
to be biased toward dustier galaxies than optically 
selected samples. 
The star-forming galaxies in our intermediate redshift 
galaxy sample deviate systematically, especially those 
with star formation rates lower than ${\rm 
\sim 10 M_{\odot} yr^{-1}}$, from the fit of Hopkins 
et al. (2001), indicating that for a given star formation 
rate, star-forming galaxies at earlier cosmic epochs 
show on average lower internal reddening than their local 
couterparts. Galaxies with star formation rate larger than
${\rm \sim 10 M_{\odot} yr^{-1}}$ display Balmer decrements 
and a scatter similar to what is observed for local 
star-forming galaxies. 

The upper panel of Fig.\ref{hahb_sfr_oh} shows the 
relationship between gas phase oxygen abundance and 
extinction-corrected star formation rate for the local 
comparative sample and the sample of intermediate redshift 
star-forming galaxies. The trend observed locally between
oxygen abundance and extinction-uncorrected star formation
rate (see Fig.\ref{oh_sfr}) is still visible after the 
extinction correction of the star formation rate. 
A similar trend has also been observed recently for 
a sample of blue compact galaxies (Kong 2004).
No convincing trend is observed for star-forming galaxies 
in our intermediate redshift sample. Metal-poor luminous 
galaxies at intermediate redshifts show higher 
extinction-corrected star formation rates than observed 
locally for galaxies with similar oxygen abundances, 
while metal-rich luminous galaxies show 
extinction-corrected star formation rates similar 
to those observed for their local counterparts in 
agreement with what was reported recently for a sample 
of Canada-France Redshift Survey galaxies (Maier et al.
2005). 

There have been contradictory claims regarding a possible
correlation between dust obscuration and metallicity of
galaxies in the local universe. Zaritsky et al. (1994)
have reported no clear evidence for a sensitivity of 
internal reddening to abundance in a sample of disk 
galaxies, suggesting that abundance and extinction are
not neccessarily linked. It has been found, however, for 
other local galaxy samples that the extinction derived 
from the Balmer decrement correlates with metallicity 
(e.g., Stasi\'nska \& Sodre 2001; Kong et al. 2002).
The observed trends between galaxy luminosity, 
metallicity, dust extinction, and star formation rate 
for local star-forming galaxies, suggest that the main 
driver of the extinction of galaxies is their mass, 
combined with their metallicity (and probably the presence 
of old stellar populations; Stasi\'nska et al. 2004). 
Fig.\ref{oh_ebmv} shows the relationship between gas 
phase oxygen abundance and colour excess for the local 
comparison sample and our sample of intermediate 
redshift galaxies. Locally, a positive trend is apparent 
between oxygen abundance and colour excess, although 
with a large scatter (see Stasi\'nska \& Sodr\'e 2001; 
Kewley et al. 2004 who analysed the properties of the 
same local galaxy sample). Strikingly, luminous 
star-forming galaxies in our intermediate redshift 
sample and local star-forming galaxies in the NGFS sample 
distribute similarly in the oxygen abundance vs. colour 
excess diagram. 
Combining this with the absence of a correlation between 
galaxy luminosity and colour excess for intermediate 
redshift luminous star-forming galaxies, one might 
conclude that the dust obscuration is primarily a function 
of the level of galaxy chemical enrichement rather than 
luminosity. This emphasizes the importance of accurate 
estimates of dust obscuration, using Balmer decrements or 
the energy balance between different observed star-formation 
indicators, to determine the oxygen abundances properly 
(see also Maier et al. 2005). Consequently, the evolution 
of the luminosity--metallicity relation as a function of 
cosmic epoch implies that one cannot use the local 
relationship between galaxy luminosity and colour excess 
to estimate the dust obscuration affecting galaxies at 
earlier cosmic epochs based only on their luminosities.

\begin{table}
\caption{Coordinates of the objects in our sample of field 
galaxies }
\label{gal_coord}
\begin{tabular}{@{}lccccccc}
\hline
   ID & R.A. (J2000) & Dec. (J2000)  \\
 \hline

1   & 02:39:57.8  & -01:33:10      \\
2   & 02:39:51.8  & -01:35:21      \\
3   & 02:39:48.1  & -01:38:16      \\
4   & 02:40:03.6  & -01:37:56      \\
5   & 02:39:54.5  & -01:35:04      \\
6   & 02:40:00.9  & -01:36:16      \\
7   & 02:40:06.7  & -01:36:55      \\
8   & 22:58:33.7  & -34:47:43      \\
9   & 22:58:41.9  & -34:47:21      \\
10  & 00:56:59.1  & -27:40:41      \\
11  & 00:56:48.2  & -27:40:13      \\
12  & 00:56:48.6  & -27:40:01      \\
13  & 00:56:58.9  & -27:43:53      \\
14  & 00:56:46.3  & -27:42:50      \\
15  & 00:56:50.9  & -27:38:01      \\
16  & 00:56:47.2  & -27:38:33      \\
17  & 00:56:55.6  & -27:39:08      \\
18  & 00:57:11.5  & -27:39:48      \\
19  & 00:57:07.9  & -27:39:28      \\
20  & 04:43:06.7  & +02:12:15      \\
21  & 04:43:14.4  & +02:10:30      \\
22  & 20:56:23.2  & -04:34:41      \\
23  & 20:56:22.4  & -04:37:49      \\
24  & 20:56:20.2  & -04:37:50      \\
25  & 20:56:22.6  & -04:41:32      \\
26  & 20:56:24.4  & -04:39:53      \\
27  & 10:57:01.2  & -03:34:20      \\
28  & 00:18:32.8  & +16:22:18      \\
29  & 00:18:27.0  & +16:26:59      \\
30  & 00:18:32.1  & +16:25:22      \\
31  & 00:18:32.8  & +16:26:10      \\
32  & 00:18:15.6  & +16:25:05      \\
33  & 00:18:17.2  & +16:25:34      \\
34  & 00:18:21.0  & +16:26:14      \\
35  & 00:18:19.2  & +16:25:44      \\
36  & 00:18:18.2  & +16:23:58      \\
37  & 00:18:37.8  & +16:24:56      \\
38  & 16:23:42.6  & +26:31:14      \\
39  & 16:23:36.0  & +26:34:17      \\
40  & 16:23:40.1  & +26:34:05      \\
41  & 20:56:19.6  & -04:38:48      \\
42  & 20:56:24.9  & -04:37:38      \\
43  & 20:56:25.2  & -04:36:00      \\
44  & 20:56:32.5  & -04:36:27      \\
 \hline
\end{tabular}
\end{table}

\setlength{\tabcolsep}{1.3pt}
\begin{table*}
 \centering
 \begin{minipage}{280mm}
\caption{\label{gal_prop}
Properties of our sample of intermediate redshift field galaxies.
The columns give the ID, redshift, absolute rest-frame $B$-band magnitude,
rest-frame equivalent widths of {\oii}, {\hbeta} and {\oiiis}, rotation
velocity, emission scalelength, star formation rate determined from {\hbeta},
and the colour excess due to internal dust extinction, respectively.}

  \begin{tabular}{@{}lccccccccccc@{}}
  \hline
ID & $z$ & $M_{B}$ & EW({\oii}) & EW({\hbeta}) & EW({\oiiis}) & $V_{rot}$ (${\rm km s^{-1}}$) & size (kpc) & {\doh} & SFR(${\rm M_{\odot} yr^{-1}}$) & E(B-V)\\
   &     & mag   & (\AA)  & (\AA)  & (\AA)  & (${\rm km s^{-1}}$) & (kpc) & & (${\rm M_{\odot} yr^{-1}})$ & (mag) \\
 \hline
1  & 0.547 &  -21.47 $\pm$ 0.11 & 12.57 $\pm$ 1.67 &  6.06 $\pm$ 0.97 &  4.67 $\pm$ 0.77 & 234.86 $\pm$ 37.75       & 5.18 $\pm$ 0.57       & 8.86 $\pm$ 0.06 & 2.62 $\pm$ 0.42 & 0.44 $\pm$ 0.25 \\
2  & 0.305 &  -21.01 $\pm$ 0.19 &  8.66 $\pm$ 0.87 &  4.36 $\pm$ 0.28 &  1.11 $\pm$ 0.19 &$98.12_{+17.42}^{-40.91}$ & 3.67 $\pm$ 0.63       & 8.93 $\pm$ 0.02 & 1.24 $\pm$ 0.08 & 0.23 $\pm$ 0.13 \\
3  & 0.325 &  -19.08 $\pm$ 0.17 & 54.38 $\pm$ 4.09 &  9.18 $\pm$ 0.99 & 10.07 $\pm$ 0.98 &  99.02 $\pm$ 11.5        & 3.43 $\pm$ 0.15       & 8.40 $\pm$ 0.15 & 0.44 $\pm$ 0.05 & 0.000 $\pm$ 0.22 \\
4  & 0.682 &  -21.59 $\pm$ 0.04 & 47.99 $\pm$ 2.24 & 15.92 $\pm$ 1.04 & 15.31 $\pm$ 1.60 & ...                      & ...                   & 8.73 $\pm$ 0.05 & 7.68 $\pm$ 0.50 & 0.000 $\pm$ 0.12 \\
5  & 0.346 &  -20.97 $\pm$ 0.15 & 16.99 $\pm$ 1.81 &  4.31 $\pm$ 0.8  &  1.59 $\pm$ 0.64 &$181.59_{+14.34}^{-18.71}$&$16.1_{+1.87}^{-0.35}$ & 8.68 $\pm$ 0.16 & 1.18 $\pm$ 0.22 & 0.09 $\pm$ 0.34  \\
6  & 0.23  &  -20.44 $\pm$ 0.03 & 36.94 $\pm$ 1.54 & 13.78 $\pm$ 0.44 &  6.66 $\pm$ 0.34 &  48.73 $\pm$ 6.3         & $2.18_{+0.16}^{-0.14}$& 8.82 $\pm$ 0.02 & 2.32 $\pm$ 0.07 & 0.07 $\pm$ 0.07  \\
7  & 0.421 &  -21.14 $\pm$ 0.08 & 47.31 $\pm$ 1.79 & 11.68 $\pm$ 0.77 & 31.55 $\pm$ 0.54 & ...                      & ...                   & 8.47 $\pm$ 0.07 & 3.73 $\pm$ 0.24 & 0.21 $\pm$ 0.12  \\
8  & 0.351 &  -20.2  $\pm$ 0.06 & 21.52 $\pm$ 2.07 &  5.90 $\pm$ 0.87 &  1.32 $\pm$ 0.75 &$170.61_{+13.57}^{-14.74}$& $4.16_{+0.38}^{-0.48}$& 8.72 $\pm$ 0.12 & 0.79 $\pm$ 0.12 & 0.000 $\pm$ 0.28 \\
9  & 0.5   &  -23.02 $\pm$ 0.04 & 54.97 $\pm$ 2.09 & 16.60 $\pm$ 1.06 & 15.78 $\pm$ 0.59 & ...                      & ...                   & 8.70 $\pm$ 0.05 &29.68 $\pm$ 1.89 & 0.000 $\pm$ 0.11 \\
10 & 0.598 &  -21.35 $\pm$ 0.13 & 15.19 $\pm$ 1.85 &  5.13 $\pm$ 1.42 & 14.53 $\pm$ 0.65 & ...                      & ...                   & 8.57 $\pm$ 0.24 & 1.99 $\pm$ 0.55 & 0.62 $\pm$ 0.46  \\
11 & 0.348 &  -20.02 $\pm$ 0.13 & 32.87 $\pm$ 2.81 &  4.61 $\pm$ 1.11 &  4.38 $\pm$ 0.74 & ...                      & ...                   &...              &       ...       & ...              \\     
12 & 0.225 &  -19.08 $\pm$ 0.18 & 55.68 $\pm$ 2.42 &  9.94 $\pm$ 0.87 & 29.67 $\pm$ 0.62 & ...                      & ...                   & ...             &        ...      &  ...            \\      
13 & 0.651 &  -22.31 $\pm$ 0.13 & 52.66 $\pm$ 0.85 & 18.99 $\pm$ 0.49 & 18.47 $\pm$ 0.31 & ...                      & ...                   & 8.76 $\pm$ 0.02 &17.71 $\pm$ 0.46 & 0.11 $\pm$ 0.04 \\
14 & 0.237 &  -21.84 $\pm$ 0.18 & 16.59 $\pm$ 1.11 &  2.42 $\pm$ 0.9  &  2.20 $\pm$ 0.26 & 228.3  $\pm$ 12.4        &11.8  $\pm$ 0.48       & ...             &    ...          &                 \\
15 & 0.71  &  -21.68 $\pm$ 0.16 & 51.89 $\pm$ 2.52 & 14.96 $\pm$ 1.00 & 12.92 $\pm$ 0.59 & 176.66 $\pm$ 41.05       & 3.16 $\pm$ 0.13       & 8.69 $\pm$ 0.05 & 7.84 $\pm$ 0.53 & 0.02 $\pm$ 0.12 \\
16 & 0.66  &  -21.18 $\pm$ 0.14 & 37.96 $\pm$ 1.85 &  9.85 $\pm$ 0.87 & 11.28 $\pm$ 1.07 & $126.28_{+9.39}^{-9.6}$  &  $3.4_{+0.16}^{-0.18}$& 8.62 $\pm$ 0.08 & 3.26 $\pm$ 0.29 & 0.03 $\pm$ 0.16 \\
17 & 0.585 &  -22.11 $\pm$ 0.12 & 20.63 $\pm$ 0.87 &  9.77 $\pm$ 0.68 &  4.55 $\pm$ 0.29 & 244.17 $\pm$ 66.07       & 4.47 $\pm$ 0.46       & 8.89 $\pm$ 0.02 & 7.59 $\pm$ 0.53 & 0.12 $\pm$ 0.10 \\
18 & 0.487 &          ...       & 38.57 $\pm$ 0.53 & 19.61 $\pm$ 0.64 & 74.54 $\pm$ 2.48 & ...                      & ...                   & 8.59 $\pm$ 0.03 &        ...      & ...             \\      
19 & 0.653 &  -21.2  $\pm$ 0.13 & 58.73 $\pm$ 1.77 & 21.14 $\pm$ 1.09 & 16.69 $\pm$ 1.39 & 155.08 $\pm$ 19.75       & $4.57_{+0.05}^{-0.14}$& 8.78 $\pm$ 0.03 & 7.13 $\pm$ 0.37 & 0.08 $\pm$ 0.09  \\
20 & 0.318 &  -21.08 $\pm$ 0.26 & 35.21 $\pm$ 2.98 &  9.75 $\pm$ 1.01 &  7.56 $\pm$ 0.81 &$168.17_{+12.6}^{-11.18}$ & $5.97_{+0.33}^{-0.28}$& 8.68 $\pm$ 0.09 & 2.95 $\pm$ 0.30 & 0.17 $\pm$ 0.20 \\
21 & 0.401 &  -20.39 $\pm$ 0.22 & 75.73 $\pm$ 7.02 & 16.94 $\pm$ 1.21 & 34.78 $\pm$ 1.58 &$114.02_{+12.4}^{-13.91}$& 2.47 $\pm$ 0.2         & 8.48 $\pm$ 0.09 & 2.72 $\pm$ 0.19 & 0.000 $\pm$ 0.17 \\
22 & 0.335 &  -20.41 $\pm$ 0.07 & 45.44 $\pm$ 3.30 & 11.97 $\pm$ 0.99 & 10.28 $\pm$ 0.82 &  $87.52_{+7.16}^{-7.72}$ & 3.46 $\pm$ 0.18       & 8.66 $\pm$ 0.08 & 1.96 $\pm$ 0.16 & 0.000 $\pm$ 0.17 \\
23 & 0.46  &  -20.53 $\pm$ 0.06 & 56.79 $\pm$ 1.66 & 16.32 $\pm$ 0.75 & 20.65 $\pm$ 0.72 & ...                      & ...                   & 8.65 $\pm$ 0.04 & 2.98 $\pm$ 0.14 & 0.01 $\pm$ 0.08 \\
24 & 0.599 &  -21.83 $\pm$ 0.12 & 12.61 $\pm$ 0.94 &  7.21 $\pm$ 0.99 &  1.14 $\pm$ 0.90 & ...                      & ...                   & 8.96 $\pm$ 0.02 & 4.33 $\pm$ 0.59 & 0.14 $\pm$ 0.17 \\
25 & 0.196 &  -19.99 $\pm$ 0.18 & 42.01 $\pm$ 1.98 &  8.19 $\pm$ 0.39 &  9.47 $\pm$ 0.33 & 125.16 $\pm$ 5.22        & 4.03 $\pm$ 0.2        & 8.49 $\pm$ 0.06 & 0.91 $\pm$ 0.04 & 0.08 $\pm$ 0.11 \\
26 & 0.335 &  -19.56 $\pm$ 0.04 & 63.67 $\pm$ 3.26 & 16.80 $\pm$ 0.96 & 19.79 $\pm$ 0.66 & ...                      & ...                   & 8.63 $\pm$ 0.05 & 1.26 $\pm$ 0.07 & 0.05 $\pm$ 0.11 \\
27 & 0.47  &  -21.62 $\pm$ 0.04 & 33.07 $\pm$ 0.95 & 12.10 $\pm$ 0.68 &  5.93 $\pm$ 0.43 & 180.38 $\pm$ 8.1         & 3.96 $\pm$ 0.4        & 8.81 $\pm$ 0.03 & 5.99 $\pm$ 0.34 & 0.000 $\pm$ 0.09 \\
28 & 0.655 &  -21.79 $\pm$ 0.08 &  3.69 $\pm$ 4.15 &  2.65 $\pm$ 1.69 &  3.30 $\pm$ 3.71 & ...                      & ...                   & 8.88 $\pm$ 0.25 & 1.53 $\pm$ 0.98 & 0.83 $\pm$ 1.45 \\
29 & 0.286 &  -20.02 $\pm$ 0.17 &  4.94 $\pm$ 1.72 &  0.77 $\pm$ 0.48 &  1.06 $\pm$ 0.81 & ...                      & ...                   & ...             &  ...            & ...             \\
30 & 0.302 &  -19.61 $\pm$ 0.06 & 23.27 $\pm$ 2.73 &  4.33 $\pm$ 3.18 &  3.12 $\pm$ 0.95 & ...                      & ...                   & 8.49 $\pm$ 0.79 & 0.34 $\pm$ 0.25 & 0.000 $\pm$ 1.26 \\
31 & 0.658 &  -21.67 $\pm$ 0.13 & 30.45 $\pm$ 0.89 & 17.50 $\pm$ 1.08 &  9.32 $\pm$ 1.16 & ...                      & ...                   & 8.92 $\pm$ 0.01 & 9.08 $\pm$ 0.56 & 0.11 $\pm$ 0.08 \\
32 & 0.447 &  -21.26 $\pm$ 0.07 & 52.39 $\pm$ 0.95 & 15.27 $\pm$ 0.62 & 25.79 $\pm$ 0.76 & ...                      & ...                   & 8.62 $\pm$ 0.03 & 5.44 $\pm$ 0.22 & 0.03 $\pm$ 0.07 \\
33 & 0.388 &  -20.16 $\pm$ 0.08 & 13.39 $\pm$ 1.09 &  8.41 $\pm$ 0.77 &  2.24 $\pm$ 0.86 & ...                      & ...                   & 8.97 $\pm$ 0.01 & 1.09 $\pm$ 0.10 & 0.14 $\pm$ 0.13 \\
34 & 0.387 &  -20.22 $\pm$ 0.07 & 52.68 $\pm$ 1.95 & 14.32 $\pm$ 1.15 & 32.74 $\pm$ 1.38 & $148.69_{+11.4}^{-9.73}$ & 4.48 $\pm$ 0.3        & 8.54 $\pm$ 0.08 & 1.97 $\pm$ 0.16 & 0.05 $\pm$ 0.14 \\
35 & 0.447 &  -20.14 $\pm$ 0.08 & 39.93 $\pm$ 2.32 & 13.68 $\pm$ 1.32 & 11.12 $\pm$ 3.14 & $210.19_{+19.7}^{-22.}$  & 2.69 $\pm$ 0.14       & 8.76 $\pm$ 0.07 & 1.75 $\pm$ 0.17 & 0.000 $\pm$ 0.18 \\
36 & 0.35  &  -20.11 $\pm$ 0.09 & 46.43 $\pm$ 1.22 & 12.96 $\pm$ 0.8  & 17.01 $\pm$ 0.5  & ...                      & ...                   & 8.64 $\pm$ 0.05 & 1.61 $\pm$ 0.10 & 0.000 $\pm$ 0.10 \\
37 & 0.397 &  -20.97 $\pm$ 0.07 & 19.89 $\pm$ 0.71 & 10.17 $\pm$ 0.66 &  3.37 $\pm$ 0.29 & ...                      & ...                   & 8.92 $\pm$ 0.01 & 2.78 $\pm$ 0.18 & 0.12 $\pm$ 0.08 \\
38 & 0.476 &  -21.51 $\pm$ 0.07 & 28.38 $\pm$ 1.19 & 10.05 $\pm$ 0.56 &  7.26 $\pm$ 0.62 & ...                      & ...                   & 8.78 $\pm$ 0.03 & 4.50 $\pm$ 0.25 & 0.09 $\pm$ 0.09 \\
39 & 0.345 &  -19.73 $\pm$ 0.06 & 21.7  $\pm$ 4.   &  4.36 $\pm$ 3.24 &  1.87 $\pm$ 0.70 & ...                      & ...                   & 8.55 $\pm$ 0.74 & 0.38 $\pm$ 0.28 & 0.15 $\pm$ 1.29 \\
40 & 0.37  &  -20.15 $\pm$ 0.09 & 39.27 $\pm$ 2.77 &  9.42 $\pm$ 1.24 & 14.77 $\pm$ 1.28 &$144.26_{+14.92}^{-17.14}$& $4.07_{+0.26}^{-0.3}$ & 8.55 $\pm$ 0.13 & 1.21 $\pm$ 0.16 & 0.24 $\pm$ 0.24  \\
41 & 0.371 &  -20.56 $\pm$ 0.05 & 22.08 $\pm$ 1.89 &  5.14 $\pm$ 1.09 &  9.08 $\pm$ 1.06 &$188.34_{+19.54}^{-19.37}$& $8.71_{+1.15}^{-0.82}$& 8.52 $\pm$ 0.21 & 0.96 $\pm$ 0.20 & 0.19 $\pm$ 0.37 \\
42 & 0.464 &  -20.32 $\pm$ 0.14 & 13.07 $\pm$ 1.86 &  6.38 $\pm$ 1.01 &  3.43 $\pm$ 1.05 & ...                      & ...                   & 8.89 $\pm$ 0.05 & 0.96 $\pm$ 0.15 & 0.13 $\pm$ 0.25 \\
43 & 0.444 &  -21.12 $\pm$ 0.2  & 27.44 $\pm$ 1.24 & 10.12 $\pm$ 0.84 &  7.73 $\pm$ 0.64 & ...                      & ...                   & 8.79 $\pm$ 0.04 & 3.17 $\pm$ 0.26 & 0.14 $\pm$ 0.13 \\
44 & 0.397 &  -20.03 $\pm$ 0.08 & 14.31 $\pm$ 2.10 &  5.46 $\pm$ 1.20 &  2.17 $\pm$ 0.84 &$195.81_{+13.33}^{-10.35}$&$4.23_{+0.31}^{-0.23}$ & 8.84 $\pm$ 0.10 & 0.63 $\pm$ 0.14 & 0.03 $\pm$ 0.35  \\

\hline
\end{tabular}
\end{minipage}
\end{table*}

\section{Discussion}
\label{disc}

We have thus found that the properties of the interstellar 
star-forming gas in massive and large field galaxies at 
intermediate redshifts cover a wide range, extending 
from those observed for local bright metal-rich galaxies, 
i.e., weak emission lines, low ionization conditions, 
high oxygen abundances and high dust obscuration, to 
those of local metal-poor dwarf galaxies which exhibit 
signatures of strong on-going star-formation activity, 
i.e., strong emission lines, high ionization conditions, 
low oxygen abundances, and low dust obscuration. 

A fraction of massive field galaxies at intermediate 
redshifts have already undergone a significant 
chemical enrichment, as indicated by their high oxygen 
abundances. This suggests that this population of 
galaxies have already formed a large fraction of the 
stellar mass observed at the present epoch in massive, 
metal-rich galaxies.
This is consistent with the reported mild evolution 
of the galaxy stellar mass function of the massive 
galaxy population since $z \sim 1$, implying that the 
evolution of these galaxies is essentially complete 
by this redshift (e.g., Brinchmann \& Ellis 2000, 
Fontana et al. 2004, Drory et al. 2004, Bundy et al. 
2005). The relationship between stellar mass and 
maximum rotation velocity, i.e., the baryonic 
Tully-Fisher relation, out to $z \sim 1$ seems to be 
largely consistent with the local relation (Conselice 
et al. 2005; but see Flores et al. 2005 for a different 
view). These observational constraints are consistent 
with the downsizing picture for galaxy evolution.
In this scenario, introduced initially by Cowie et al. 
(1996), the mass threshold below which star-forming 
galaxies, defined as galaxies in which the formation 
timescale is less than the Hubble time at the galaxy 
redshifts, are found, decreases with cosmic epoch. 
Consequently, the star formation activity, and hence 
chemical evolution, seems to stop within the highest 
mass galaxies at earlier cosmic epochs than for less 
massive galaxies (see also Fig. 5 of Maier et al. 2006).

On the other hand, a sub-population of massive and 
luminous intermediate redshift star-forming galaxies,  
with emission length sizes larger than 2 kpc, appear 
to be at earlier stages of the assembly of their content 
of metals, as indicated by their low oxygen abundances. 
This suggests that they are probably also at earlier 
stages of assembling their stellar masses, as both 
stellar and metal contents of galaxies are thought to 
be linked. Metal-poor luminous galaxies at intermediate
redshift are therefore expected to deviate from the 
local baryonic Tully-Fisher relation.

Hammer et al. (2005) have presented a body of evidence to 
show that present-day intermediate stellar mass galaxies, 
i.e., with stellar masses of 
$3\,$--${\,\rm 30 \times 10^{10} M_{\odot}}$, have built 
up half of their stellar mass since $z \sim 1$, and 
emphasize the importance of luminous infrared galaxies, 
thought to be starbursts resulting from merging at these 
masses.
The star formation in luminous infrared galaxies accounts 
for most of the stellar mass growth in these galaxies 
since $z \sim 1$. They argue that recent merging and 
gas infall explain both star formation history and 
morphological changes in a hierarchical galaxy formation 
scheme, where more than 50 per cent of spiral galaxies 
experienced their last major merger event less than 8~Gyr 
ago. Intermediate redshift luminous infrared galaxies 
show oxygen abundances two times lower than local bright 
disk galaxies (Liang et al. 2004). 
The presence of a population of intermediate redshift 
massive and luminous galaxies with oxygen abundances 
lower than those observed locally for similar galaxies, 
supports the scenario in which the assembly of intermediate 
stellar mass galaxies is still operating between $z \sim 1$ 
and $z = 0$.

The sample studied in this paper is by no means complete. 
It is not possible therefore to assess the implication 
of our findings for constraining galaxy assembly from 
$z \sim 1$ to the present epoch.  Additional work to 
determine the properties of metal-poor massive galaxies 
and their evolution, e.g., number density, stellar masses, 
morphologies, and stellar mass function, is needed to help 
reveal the nature of star formation during the last 
$\sim\,8$ Gyr. In the near future it will be possible to 
conduct further studies to measure chemical properties 
and stellar masses of star-forming galaxies at intermediate 
redshifts, in order to tightly constrain the assembly of 
the stellar and metal content of present day galaxies. 
Large galaxy surveys like DEEP2 (Davis et al. 2003) and 
EDisCS (White et al. 2005) are promising, as they will 
observe large samples of galaxies distributed over a wide 
area and redshift range.

\section{Summary \& Conclusions}
\label{concl}

We have used spectrophotometric data for a sample of 
luminous ($M_{B} \la -19$), mostly disk, field galaxies 
in the redshift range $0.2 \la z \la 0.8$, with measured 
maximum rotation velocities and emission scale lengths, 
to investigate the properties of the interstellar 
emitting gas of massive galaxies at intermediate 
redshift.

Emission line equivalent widths and excitation- and 
metallicity-sensitive emission line diagnostic ratios 
of the massive star-forming galaxies in our sample cover 
similar ranges to those observed for local emission line 
galaxies over a wide range of luminosities, i.e., 
$-14.0 \la M_{B} \la -22.0$. The properties of the 
interstellar emitting gas for a subsample of massive and 
luminous field galaxies at intermediate redshifts are 
similar to those observed for faint and metal-poor galaxies, 
with moderate excitation-sensitive diagnostic ratios, at 
the present cosmic epoch.

The oxygen abundance of interstellar emitting gas has been 
estimated using the ``strong-line'' method. These estimated 
oxygen abundances range from 8.4 to 9.0 in units of {\doh}. 
Our sample galaxies exhibit a luminosity--metallicity 
relation different from that of local galaxies. A subsample 
of massive galaxies show oxygen abundances that are 
consistent with what is observed locally for their local 
couterparts. However, a fraction of massive star-forming 
galaxies in our sample have low oxygen abundances that are 
observed locally only for much fainter galaxies, in agreement 
with the findings of other recent investigations of the 
chemical properties of star-forming galaxies at intermediate 
redshifts (Kobulnicky et al. 2003; Liang et al. 2004, Maier 
et al. 2005). Oxygen abundances are not found to correlated 
with either the maximum rotation velocity nor with the 
emission scale length size of the galaxy. 

Luminous and large field galaxies at intermediate redshifts 
show similar extinction-uncorrected and intrinsic star 
formation rates to their local couterparts, independent
of their gas phase oxygen abundances. The nebular extinction 
as derived from the ratio of the extinction-uncorrected star 
formation rates based on {\oii} and {\hbeta}, respectively,
is found to span a similar range to that for star-forming 
galaxies at the present epoch, but has a lower mean than 
is observed locally for optically-selected galaxy samples. 
Intermediate redshift luminous metal-rich galaxies exhibit
similar internal reddening and intrinsic star formation 
rate to local bright metal-rich galaxies, while luminous 
metal-poor galaxies show lower internal reddening, 
similar to what is observed locally for faint metal-poor 
galaxies, but similar intrinsic star formation rate to
metal-rich galaxies.

The relationship between gas phase oxygen abundance and 
colour excess for intermediate redshift galaxies is 
similar to that observed at the present epoch. The best 
interpretation for this is that the dust obscuration is 
regulated more by galaxy metal content than luminosity.

The diversity of the properties of massive and large 
galaxies at intermediate redshifts supports the scenario
whereby galaxies are still assembling their baryonic
content between $z \sim 1$ and $z = 0$. 
Intermediate redshift massive and large galaxies with 
low gas phase oxygen abundances are most likely immature 
galaxies that will increase their metallicities and their 
stellar masses to the present epoch. 

\section*{Acknowledgments}
This work was based on observations made with ESO Telescopes 
at Paranal Observatory under programme IDs 066.A-0376 and 
069.A-0312, and on observations made with the NASA/ESA 
\emph{Hubble Space Telescope}, obtained from the data
archive at the Space Telescope Institute. STScI is operated 
by the association of Universities for Research in Astronomy, 
Inc. under the NASA contract NAS 5-26555.


\bsp

\label{lastpage}

\end{document}